\documentclass[%
rsi,
 reprint,%
 floatfix,
superscriptaddress
]{revtex4-1}

\usepackage{graphicx}% Include figure files
\usepackage{dcolumn}% Align table columns on decimal point
\usepackage{bm}% bold math
\usepackage[utf8]{inputenc}
\usepackage[T1]{fontenc}
\usepackage{mathptmx}
\usepackage{etoolbox}

\makeatletter
\def\@email#1#2{%
 \endgroup
 \patchcmd{\titleblock@produce}
  {\frontmatter@RRAPformat}
  {\frontmatter@RRAPformat{\produce@RRAP{*#1\href{mailto:#2}{#2}}}\frontmatter@RRAPformat}
  {}{}
}%
\makeatother
\begin{document}

\preprint{AIP/123-QED}

\title[A Cavity Load Lock Apparatus for Next-Generation Quantum Optics Experiments]{A Cavity Load Lock Apparatus for Next-Generation Quantum Optics Experiments}
% Force line breaks with \\
\author{Chuan Yin}
\affiliation{The Department of Physics and The James Franck Institute, University of Chicago, Chicago, IL}
\author{Henry Ando}
\affiliation{The Department of Physics and The James Franck Institute, University of Chicago, Chicago, IL}
\author{Mark Stone}
\affiliation{The Department of Physics and The James Franck Institute, University of Chicago, Chicago, IL}
\author{Danial Shadmany}
\affiliation{The Department of Physics, Stanford University, Stanford, CA}
\author{Anna Soper}
\affiliation{The Department of Applied Physics, Stanford University, Stanford, CA}
\author{Matt Jaffe}
\affiliation{The Department of Physics and The James Franck Institute, University of Chicago, Chicago, IL}
\author{Aishwarya Kumar}
\affiliation{The Department of Physics and The James Franck Institute, University of Chicago, Chicago, IL}
\author{Jonathan Simon}
\affiliation{The Department of Physics and The James Franck Institute, University of Chicago, Chicago, IL}
\affiliation{The Department of Physics, Stanford University, Stanford, CA}
\affiliation{The Department of Applied Physics, Stanford University, Stanford, CA}

 \email{chuanyin@uchicago.edu}

%\homepage{http://www.Second.institution.edu/~Charlie.Author.}

\date{\today}% It is always \today, today,

\begin{abstract}
Cavity quantum electrodynamics (QED), the study of the interaction between quantized emitters and photons confined in an optical cavity, is an important tool for quantum science in computing~\cite{pellizzari1995decoherence}, networking~\cite{mabuchi2001quantum,kimble2008quantum}, and synthetic matter~\cite{schine2016synthetic,chang2018colloquium,clark2020observation}.
In atomic cavity QED, this approach typically relies upon an ultra-high vacuum chamber that hosts a cold trapped atomic ensemble and an optical cavity. Upgrading the cavity necessitates a months-long laborious process of removing external optics, venting, replacing the resonator, baking, and replacing optics, constituting a substantial bottleneck to innovation in resonator design. In this work, we demonstrate that the flexibility of optical cavities, and the quick turnaround time in switching between them, can be restored with the vacuum loadlock technique--reducing the cycle time to install a cavity, bake it, and transport it into the science chamber to days, achieving $3\times 10^{-10}$ Torr pressure in the science chamber. By reducing vacuum limitations, this approach is particularly powerful for labs interested in quickly exploring novel optic cavities, or any other atomic physics relying on in-vacuum optics.
\end{abstract}

\maketitle

% \begin{quotation}
% The ``lead paragraph'' is encapsulated with the \LaTeX\ 
% \verb+quotation+ environment and is formatted as a single paragraph before the first section heading. 
% (The \verb+quotation+ environment reverts to its usual meaning after the first sectioning command.) 
% Note that numbered references are allowed in the lead paragraph.
% %
% The lead paragraph will only be found in an article being prepared for the journal \textit{Chaos}.
% \end{quotation}

\section{\label{sec:level1}Introduction:}
Cavity QED is useful for studying fundamental aspects of quantum mechanics such as decoherence, entanglement and measurement~\cite{mabuchi2002cavity}. Over the past decade, the AMO community has made substantial advances on exerting control over individual trapped atoms~\cite{wilk2010entanglement, schlosser2001sub, isenhower2010demonstration}, ensemble of atoms~\cite{lukin2003colloquium, ebert2015coherence, zhang2012coherent}, ultra-cold atoms in the BEC state~\cite{morsch2006dynamics, fetter2009rotating}, and atoms trapped in tweezer arrays~\cite{kaufman2021quantum, bernien2017probing, kaufman2014two}. This has interesting prospects for quantum information processing~\cite{levine2018high, xia2015randomized}, secure communication~\cite{duan2000quantum}, simulation of quantum systems~\cite{buluta2009quantum, lewenstein2007ultracold, jaksch2005cold}, and precision sensing~\cite{kurizki2015quantum}. 

Using cavities in vacuum enriches the scientific efforts of generating entangled states~\cite{chen2015carving}, studying decoherence~\cite{vaneecloo2022intracavity}, computing with Rydberg arrays~\cite{ramette2022any}, and building robust quantum networks at telecommunication wavelengths~\cite{ritter2012elementary}. Different applications have motivated a great variety of cavity designs, including twisted cavities~\cite{schine2016synthetic}, near concentric cavities~\cite{chen2022high, utama2021coupling}, fiber cavities~\cite{hunger2010fiber, brekenfeld2020quantum}, off-axis parabolic cavities~\cite{wang2019experimental}, and lens cavities~\cite{jaffe2021aberrated} have been demonstrated, but there remain exciting opportunities to explore new geometries, aberration corrected cavities~\cite{jaffe2021aberrated}, coupled cavities~\cite{stone2021optical} for atomic physics and quantum information applications. 

Integrating and testing exotic cavity geometries with cold atoms is challenging because achieving sufficient vacuum is a months-long process requiring removal of external optics, replacement of in-vacuum optics, baking of the chamber, and re-alignment of external optics. 
% One could always replace cavities by directly opening the main experimental chamber, but that necessitates a complex installation and bake-out procedure, involving removing all optics near the vacuum system, surrounding the whole setup by heating tapes, thermocouples and aluminum foils, and baking the apparatus at high temperature for weeks, in total costing us a few months of scientific downtime. 
One way around this is cryo-pumping as demonstrated in recent microwave/mmwave to optical transduction experiments~\cite{mckenna2020cryogenic,kumar2022quantum} and cryogenic cavity optomechanics~\cite{groblacher2009demonstration, falke2014strontium}, where all material-constraints are relaxed and bakes are obviated by cryo-pumping. Other methods include putting the cavity \emph{outside} of the vacuum chamber~\cite{yang2016efficient,yang2021single} to remove the need for in-vacuum optics all-together, and using loadlocks~\cite{kirby1993situ,springate2008artemis,mestres2015long,zhong2017millikelvin} for rapid cycling. The first option has limitations such as requiring much space for cryogenic technical supplies, needing vibration isolation schemes, differential contraction of materials, and poor piezo performance. The second option limits the cavity geometry to be larger than the vacuum system, increasing the mode waist \& reducing the finesse, and thus limiting the achievable single-atom cooperativity~\cite{tanji2011interaction}

The loadlock vacuum architecture employs two chambers separated by a gate valve, so venting the auxiliary chamber to air for sample installation need not disrupt the UHV environment of--or optical alignment to--the main chamber. It is a powerful approach used broadly across chemistry such as molecular beam epitaxy~\cite{hamm1994compact}~\cite{celinski2001molecular}, scanning electron microscopes~\cite{zhang2011load}, and cold atom experiments~\cite{low2007apparatus, leonard2017supersolid}, for getting a sample from atmosphere into UHV quickly. The loadlock architecture introduces substantial added complexity in the form of additional space taken by the loadlock chamber and a translator used for moving samples around, and trickier wire management, but provides otherwise seamless integration with standard cold atom technologies. 

Here we present a loadlock apparatus designed to optimize the turnaround time for cold atom cavity QED experiments.
Our apparatus consists of a science chamber and a loadlock chamber separated by a gate valve, featuring in-vacuum MOT optics, $^{87}$Rb dispensers, and a cavity carrier attached to a position manipulator. It supports a sizable cavity structure with maximal optical access to the atoms and exterior laser beams, and rapid cycling when introducing a new cavity. 
We characterize its performance by loading and transporting cold ensembles of $^{87}$Rb atoms at temperature 6$\mu$K, over 10 cm, in 30 ms. The apparatus offers a 4''x4''x3'' volume for custom cavity structures, with 40 electrical feedthroughs on plug-and-go connectors for cavity electrodes, MOT quadrupole coils, dispensers, and heating wires.

\section{Loadlock Architecture and Workflow}
\begin{figure}[ht]

		%\centering
		\includegraphics[width=.48\textwidth]{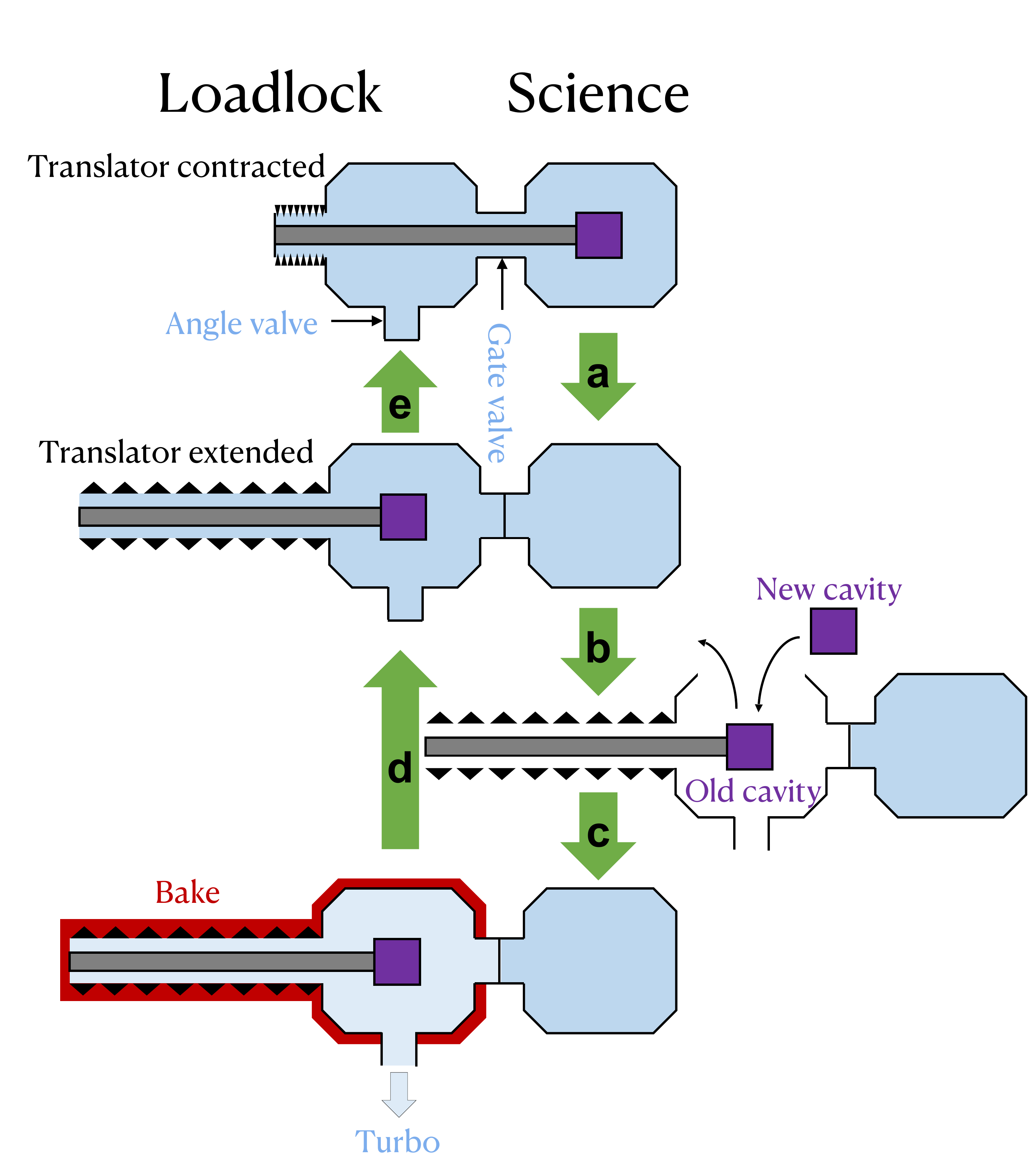}
		\caption{\textbf{A typical cycle of loading a new cavity into UHV environment}. \textbf{a.} Extend the translator to move the cavity into the loadlock chamber, close the gate valve. \textbf{b.} Vent only the loadlock chamber through the angle valve, open two flanges of the loadlock chamber, install a new cavity. \textbf{c.} Close up the loadlock chamber, pump and bake only the loadlock chamber. \textbf{d.} Shut the angle valve, cool down the loadlock chamber. \textbf{e.} Open the gate valve, contract the translator to move the cavity into the science chamber. The whole process takes about a week.}
        \label{fig:schematics_of_the_loading_process}
\end{figure}

\begin{figure*}
		%\centering
		\includegraphics[width=\textwidth]{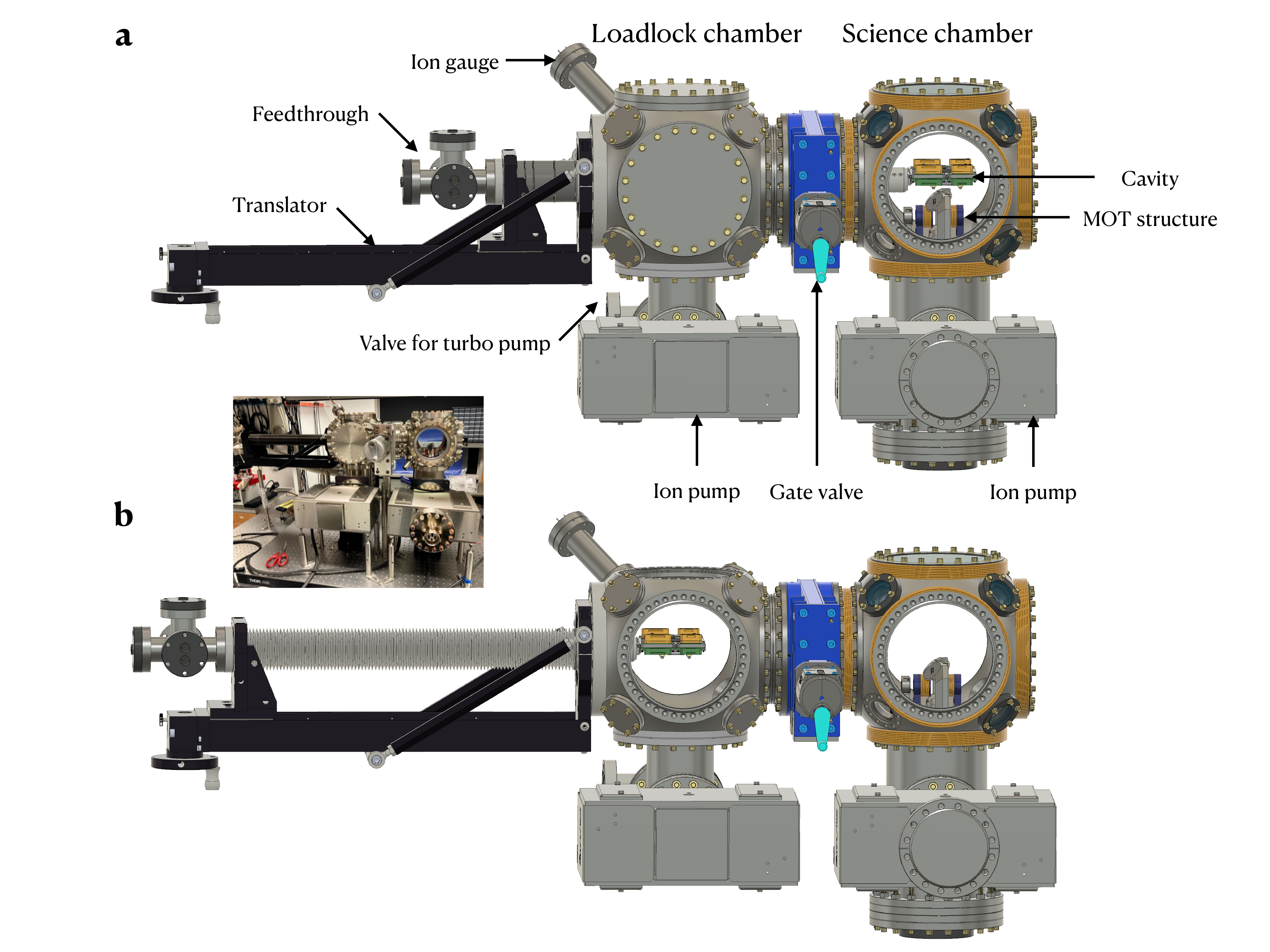}
		\caption{\textbf{Schematic of the load lock apparatus.} The system consists of two vacuum chambers: a loadlock  chamber separated from a science chamber by a gate valve. The science chamber is kept at UHV by an ion pump, and is designed to have maximal optical access. An in-vacuum MOT setup stays in the science chamber. A translator moves the carrier structure supporting an optical cavity and atomic dispensers across the two chambers. The apparatus is depicted in two configurations: \textbf{a,} science mode, where the translator is extended and the cavity is in the science chamber, and \textbf{b,} installation mode, where the translator is retracted and the cavity in the loadlock chamber. \textbf{Inset: a photograph of the apparatus.} The apparatus has dimensions 3.5’$\times$4’$\times$2.5’, and comfortably fits atop a standard 4'x8' vibration isolated optical table. Space is available under and around the vacuum chamber to conveniently set up optical breadboards. The viewports of the main chamber are AR coated to transmit multiple wavelengths of interest. The gate valve is manually actuated.}
        \label{fig:big_schematic}
\end{figure*}

%Here we describe an apparatus with fast turnaround in testing optical cavities and interfacing them with cold atoms.
To minimize the disruption of replacing the cavity on the experiment, we employ a two-chamber architecture separated by a gate valve: a main ``science'' chamber housing the UHV-clean setup for cold trapped atoms, and an auxiliary ``loadlock'' chamber where an initially unbaked cavity is installed into and independently baked clean, photos shown in Fig.~\ref{fig:big_schematic}. A z-manipulator with 60'' throw is used to move the cleaned test cavity between the loadlock chamber (see Fig.~\ref{fig:big_schematic}b) and the science chamber (see Fig.~\ref{fig:big_schematic}a). To continuously sustain the UHV-clean environment in the science chamber, an ion pump is attached to the science chamber through a vacuum cross at the bottom; to facilitate the frequent pumping and baking cycle of the loadlock chamber from the atmospheric pressure to the UHV, a turbo pump and an ion pump are attached to the loadlock chamber through a vacuum tee at the bottom, and an ion gauge at the top. 

The workflow for readying a new cavity having been exposed to air into the UHV-clean science chamber to interface with cold trapped atoms follows the steps shown in Fig.~\ref{fig:schematics_of_the_loading_process}. We first retract the translator to move the old cavity into the loadlock chamber, and close the gate valve to isolate the two chambers. e then vent only the loadlock chamber through the angle valve, open two flanges of the loadlock chamber, switch in a new cavity. We then close up the loadlock chamber, pump only the loadlock chamber with the turbo pump and an ion pump, while baking it to  $110\,^{\circ}$ C limited by in-vacuum Torr-seal glue. Once the loadlock pressure at high temperature is satisfactory, we then shut the angle valve, cool down the loadlock chamber to room temperature. Finally, if the loadlock pressure at room temperature is satisfactory, we open the gate valve, and extend the translator to move the cavity into the science chamber. 

Thanks to the loadlock architecture and the gate valve in particular, the entire process above does not expose the UHV-clean science chamber to air or heat, and therefore there is no need to remove the optics surrounding the science chamber, so the downtime required to ready a new cavity is drastically reduced to about a week. This iterative process is further simplified and expedited by our decisions to: 1) use metal flanges without viewports on the loadlock chamber to tolerate large thermal gradients, 2) apply more aggressive temperature ramps during bakeouts, 3) attach heating tapes and aluminum foil to the bottom of the loadlock chamber permanently, 4) route electrical feedthrough wires inside the tube of the translator in an organized manner, and 5) design the cavity-carrying platform to be plug-and-go style.

%%%%%%%%%%%%%%%%%%%%%%%%%%%%%%%%%%%%%%%%%%%%%%%%%%%%%%%%%%%%%%
%%%%%%%%%%%%%%%%%%%%%%%%%%%%%%%%%%%%%%%%%%%%%%%%%%%%%%%%%%%%%%
\section{In-Vacuum MOT Setup}
%%%%%%%%%%%%%%%%%%%%%%%%%%%%%%%%%%%%%%%%%%%%%%%%%%%%%%%%%%%%%%
%%%%%%%%%%%%%%%%%%%%%%%%%%%%%%%%%%%%%%%%%%%%%%%%%%%%%%%%%%%%%%

\begin{figure}[ht]
		%\centering
		\includegraphics[width=.48\textwidth]{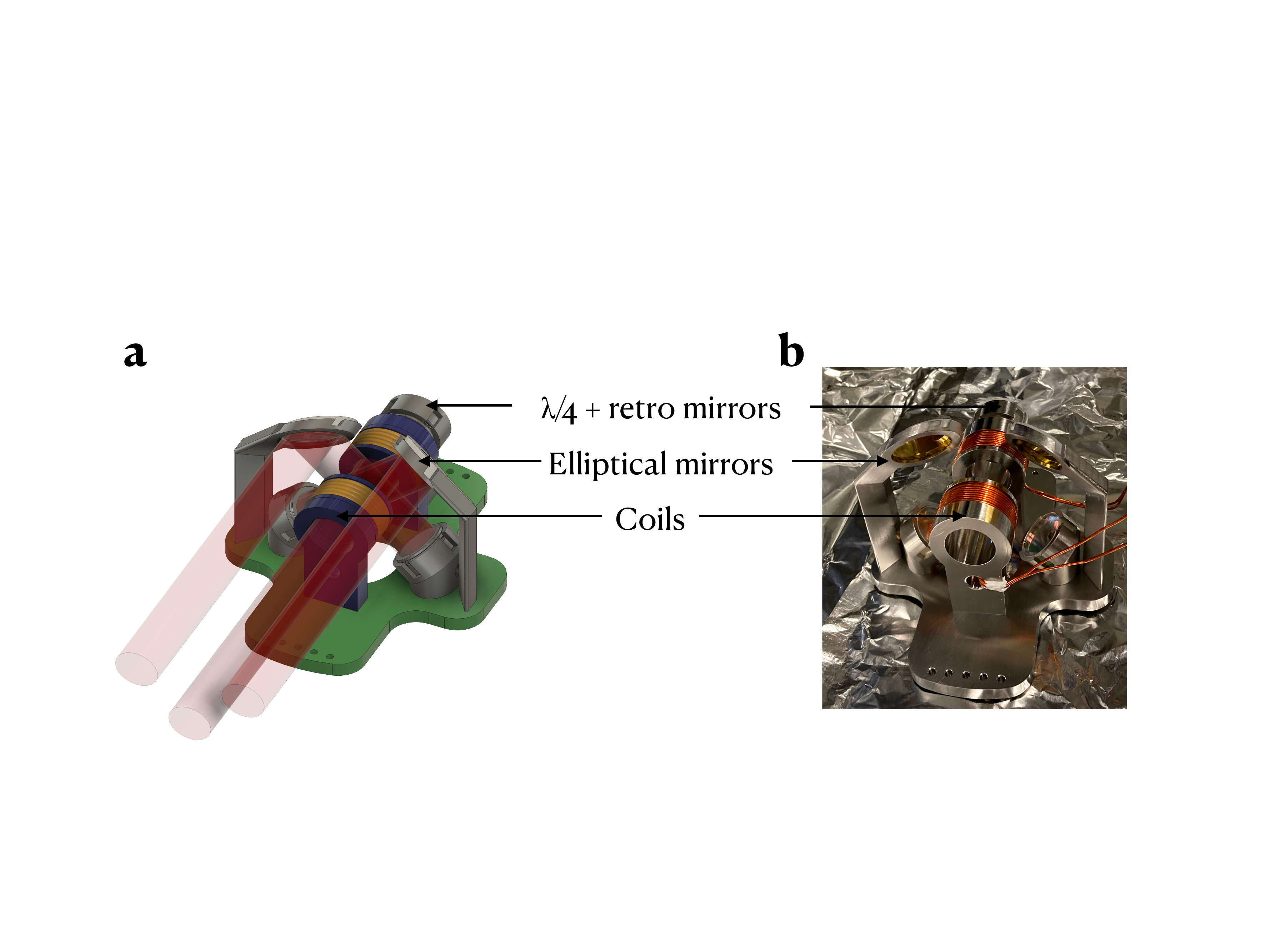}
		\caption{\textbf{In-vacuum MOT setup}. The setup, depicted schematically in \textbf{a,} and by photograph in \textbf{b,} consists of a pair of \textbf{MOT coils} for trapping atoms, and \textbf{MOT optics} for cooling atoms. The in-vacuum MOT coils are made of several layers of kapton-insulated copper wires tightly wrapped around two steel holders spaced 1’’ apart. The in-vacuum geometry allows the coils to generate a large magnetic field gradient (about 20 G/cm) with only 3 A of current. The in-vacuum design speeds up the experimental sequence by preventing eddy currents from being generated in the copper gaskets. The in-vacuum MOT optics are an assembly of steel structures holding directional mirrors and quarter waveplates necessary to achieve a MOT. All three parallel beams entering through a single viewport intersect orthogonally at the MOT location, which is below the center of the science chamber to maximize available space for the cavity structure.}
        \label{fig:MOT_setup}
\end{figure}

To maximize optical access, the science chamber is covered with four 8'' viewports orthographically, one 4.5'' viewport below through a vacuum cross, and seven 2.75'' viewports diagonally, all of which are broadband coated. At the bottom of the science chamber is located an in-vacuum MOT setup, comprising a structure for supporting directional optics for atom cooling (MOT optics), and a structure for supporting quadrupole coils for atom trapping (MOT coils). 

The in-vacuum MOT optics are mounted to a steel structure supporting circular and elliptical mirrors and quarter waveplates to direct three parallel incoming beams entering through a single viewport to orthogonally intersect at the MOT location. The MOT is formed 2" below the geometric center of the vacuum chamber to maximize the available volume for the cavity structure, which is ultimately limited to 4''x4''x3''. 

The in-vacuum MOT coils are made of 4 layers of 9 turns of kapton-insulated wires tightly wrapped around two steel holders spaced by 1’’. Placing coils inside vacuum minimizes eddy currents induced by coil turn on/off, increasing the experimental repetition rate. Additionally, since the geometry of the coils is not constrained by the (substantial) size of the vacuum chamber, a large magnetic field gradient (20 G/cm) can be generated with low electrical current (3 A) and minimal heat dissipation.

%%%%%%%%%%%%%%%%%%%%%%%%%%%%%%%%%%%%%%%%%%%%%%%%%%%%%%%%%%%%%%
%%%%%%%%%%%%%%%%%%%%%%%%%%%%%%%%%%%%%%%%%%%%%%%%%%%%%%%%%%%%%%
\section{Cavity Carrier}
%%%%%%%%%%%%%%%%%%%%%%%%%%%%%%%%%%%%%%%%%%%%%%%%%%%%%%%%%%%%%%
%%%%%%%%%%%%%%%%%%%%%%%%%%%%%%%%%%%%%%%%%%%%%%%%%%%%%%%%%%%%%%

\begin{figure}[t]
		%\centering
	\includegraphics[width=.48\textwidth]{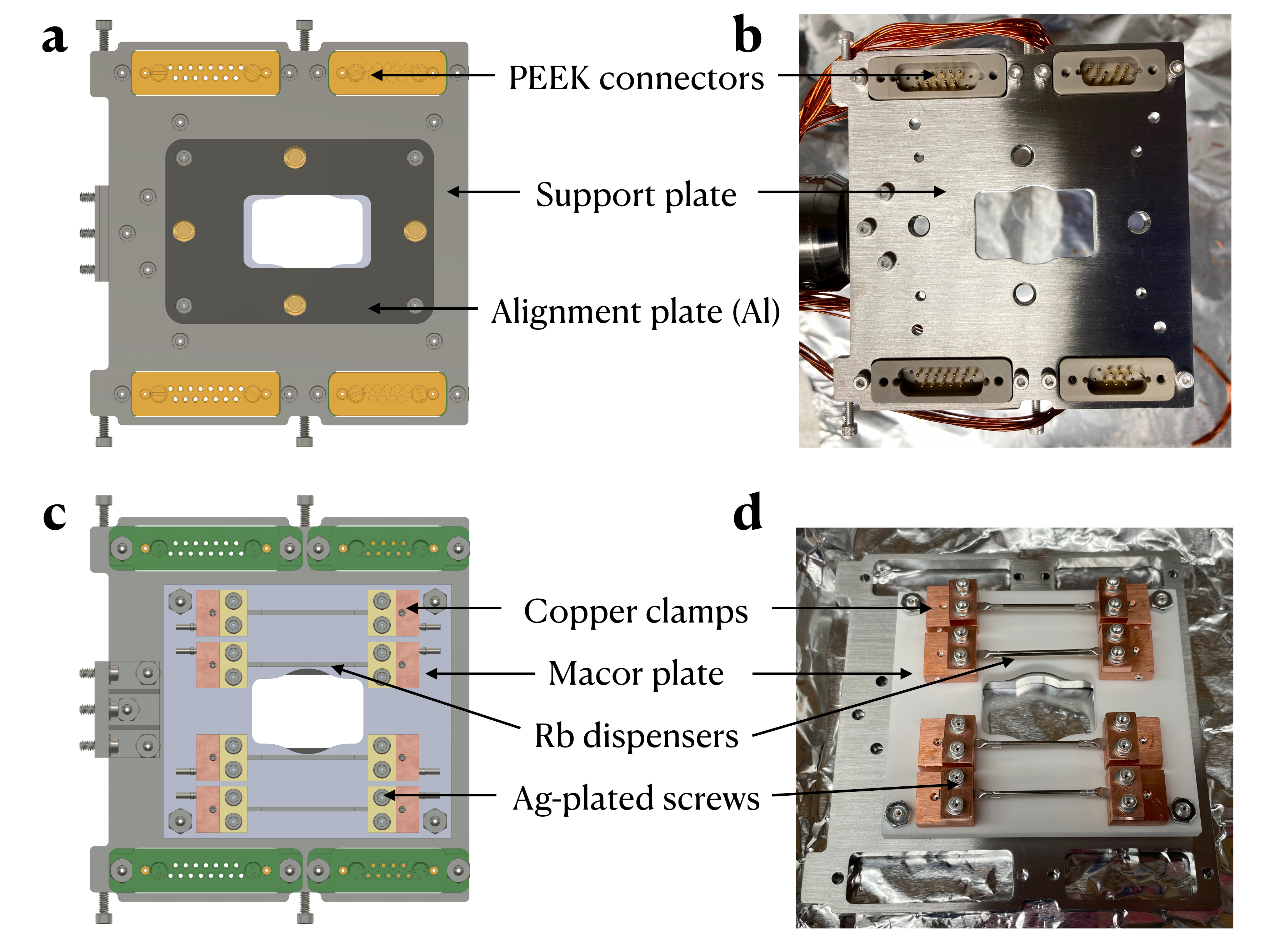}
	\caption{\textbf{Cavity carrier}. \textbf{a, b,} Looking from above, this structure carries an optical cavity above, and is secured to the end of the translator. Four dowel pins and four silver-plated screws are used to precisely align and secure the optical cavity with the carrier plate. Four pairs of PEEK connectors deliver electrical connections to the optical cavity through kapton-coated feedthrough wires bundled inside the translator tube. \textbf{c, d,} Below the cavity carrier reside four Rb dispensers which are independently controlled and easily replaceable, and each has line-of-sight access to the position of the MOT, but no line-of-sight access to the optical cavity above that can cause atom deposition on the optical surfaces. A macor plate is used as an electrical insulator to set a voltage difference across the dispensers. Silver-plated vented screws and spring washers are used to sandwich the dispensers between two layers of copper blocks, and gold pins are pressed from the side of the copper block to make secure electrical connections, in a way that the electrical connections can withstand thermal cycles of the vacuum bakeouts.}
        \label{fig:cavity_carrier}
\end{figure}
To simplify the process of moving the testing cavity between the two chambers, a cavity carrier structure connected to the end plate of the translator tube is designed to support a cavity on top with maximal flexibility (Fig.~\ref{fig:cavity_carrier}). The cavity carrier assembly consists of a central stainless steel plate which connects to the translator and structural support. On top of the cavity carrier, another smaller plate is mounted to mechanically support a cavity with four screws and four dowel pins for positional accuracy. On two sides of the central stainless plate, four female PEEK connectors are fastened below to hold a total of 40 gold pins each individually connected to a kapton-insulated wire, which are then bundled and fed inside the translator tube to the feedthrough flange. The test cavity is intended to connect via PEEK connectors to electrodes, piezo control, heating, and other electrical components.%This minimizes the turnaround time of installing a new cavity because the work is reduced to only fastening screws and dowel pins, and plugging in PEEK connectors.

%%%%%%%%%%%%%%%%%%%%%%%%%%%%%%%%%%%%%%%%%%%%%%%%%%%%%%%%%%%%%%
%%%%%%%%%%%%%%%%%%%%%%%%%%%%%%%%%%%%%%%%%%%%%%%%%%%%%%%%%%%%%%
\section{Atomic Transport System}
%%%%%%%%%%%%%%%%%%%%%%%%%%%%%%%%%%%%%%%%%%%%%%%%%%%%%%%%%%%%%%
%%%%%%%%%%%%%%%%%%%%%%%%%%%%%%%%%%%%%%%%%%%%%%%%%%%%%%%%%%%%%%
\begin{figure}[ht]
		%\centering
		\includegraphics[width=.48\textwidth]{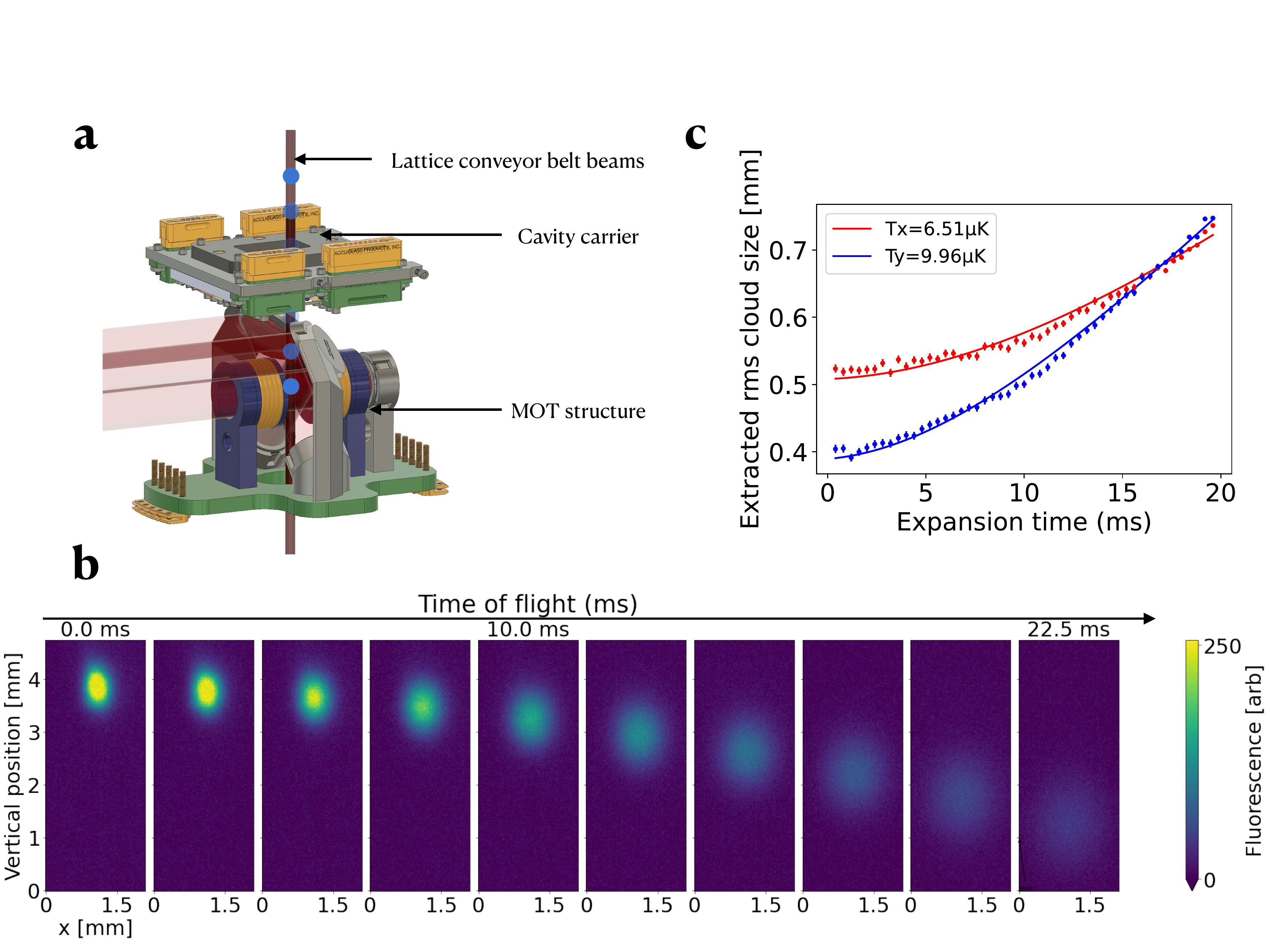}
		\caption{\textbf{Atomic transport system}. \textbf{a,} This system transports cold trapped Rb atoms from the position of the MOT upwards into the center of the optical cavity, for a total distance of about 4’’. It consists of two counter-propagating 785 nm beams, each delivering 120 mW from above or below the science chamber. The beams are frequency-shifted via two AOMs to have a relative offset up to $2$~MHz (corresponding to $\sim 0.8$m/s maximum transport speed). They are focused at the center of the transport path to create a 120 $\mu$m waist (lattice depth  about 38 $\mu$K), and path-length balanced to minimize phase-noise induced atom heating~\cite{gehm1998dynamics}.
  \textbf{Characterization of the atom source}. \textbf{b,} Upon achieving a good vacuum quality, we have cooled and trapped a $^{87}$Rb MOT using polarization gradient cooling, and performed a time-of-flight expansion measurement based on atom florescence. \textbf{c,} The spot size of the MOT vs. expansion time was plotted to extract the post-PGC temperature of the atomic cloud to be less than $10 \mu$K in both axes of the imaging plane.
  }
        \label{fig:atomic_conveyer_system}
\end{figure}

The atom source consists of four Rb dispensers held below the central cavity plate facing downwards, shown in Fig.~\ref{fig:cavity_carrier}, so the ejected atomic vapor has line-of-sight access to the center of the MOT assembly, but no line-of-sight access to the optical cavity above the plate; this minimizes deposition of Rb atoms on the cavity mirrors. The dispensers are part of the transportable setup for easy replacement, and can be individually controlled. Structurally, the dispensers are sandwiched between copper blocks which are tensioned by silver-plated, vented screws and spring washers \footnote{The spring washers are important because we have, in the past, observed that the time-varying thermal load associated with turning the dispensers on and off can eventually loosen the screws and break electrical contact}, and electrically isolated by a macor plate suspended below the central stainless plate.

Once an atomic cloud is formed by the Rb dispensers (Fig.~\ref{fig:atomic_conveyer_system}), cooled by the MOT lasers (Fig.~\ref{fig:MOT_setup}), and trapped by the magnetic field gradient (Fig.~\ref{fig:MOT_setup}) at near the bottom of the science chamber, we transport the atomic cloud upwards by 4'' to the testing cavity via an optical conveyor belt. 
The optical conveyor belt is made of two counter-propagating 120 mW, 785 nm beams optically amplified from an Innovative Photonics Solutions laser with 100 kHz linewidth, and with the differential optical path length balanced to minimize atom heating due to laser phase noise.
The lattice is translated by inducing an adjustable frequency offset up to $2$ MHz between the beams with two double-pass acousto-optic modulators, typically reaching a maximum atomic acceleration of $1500$~m/s$^2$. 
Each beam is focused at the center of the transport path to a waist of 120 $\mu$m, resulting in an axial trap frequency of $110$ kHz and a trap depth of $6$ MHz.

After the initial construction, the pressure of the science chamber was found to be $3\times 10^{-10}$ Torr. After we installed the first cavity into the loadlock chamber and transported it into the science chamber, the pressure of the science chamber was $4\times 10^{-10}$ Torr, and the final loadlock pressure was $4\times 10^{-10}$ Torr after opening the gate valve and transporting the cavity. We have cooled and trapped a $^{87}$Rb MOT with post-PGC temperature of $\sim 10 \mu$K, as determined by time-of-flight expansion measurement, shown in Fig.~\ref{fig:atomic_conveyer_system}. This figure shows the florescence imaging with a FLIR CCD camera, a fast triggerable imaging camera, and a fitted plot.

%%%%%%%%%%%%%%%%%%%%%%%%%%%%%%%%%%%%%%%%%%%%%%%%%%%%%%%%%%%%%%
%%%%%%%%%%%%%%%%%%%%%%%%%%%%%%%%%%%%%%%%%%%%%%%%%%%%%%%%%%%%%%
\section{Outlook}
%%%%%%%%%%%%%%%%%%%%%%%%%%%%%%%%%%%%%%%%%%%%%%%%%%%%%%%%%%%%%%
%%%%%%%%%%%%%%%%%%%%%%%%%%%%%%%%%%%%%%%%%%%%%%%%%%%%%%%%%%%%%%
We have installed the first cavity into the loadlock chamber, pumped and baked and transferred to the science chamber. At the time of the transfer, the loadlock chamber's ion pump read $4.5\times 10^{-10}$ Torr, and the science chamber's ion pump read $4\times 10^{-10}$ Torr, which established a UHV status successfully. We anticipate numerous upcoming opportunities to harness this technology to benchmark novel resonators coupled to atoms~\cite{jaffe2021aberrated,jaffe2022understanding,schine2016synthetic} for many-body physics~\cite{georgakopoulos2018theory}. %expect more cavities in future to go through a similar process.

\begin{acknowledgments}
This work was supported primarily by AFOSR DURIP FA9550-19-1-0140, AFOSR MURI FA9550-19-1-0399, and AFOSR grant FA9550-22-1-0279. This work was also supported by the University of Chicago Materials Research Science and Engineering Center, which is funded by the National Science Foundation under award number DMR-1420709. Additional funding for Henry Ando was provided by the National Science Foundation Graduate Research Fellowship under Grant No. DGE 1746045.
\end{acknowledgments}

\nocite{*}
\bibliography{loadlock}% Produces the bibliography via BibTeX.

\begin{thebibliography}{73}%
\makeatletter
\providecommand \@ifxundefined [1]{%
 \@ifx{#1\undefined}
}%
\providecommand \@ifnum [1]{%
 \ifnum #1\expandafter \@firstoftwo
 \else \expandafter \@secondoftwo
 \fi
}%
\providecommand \@ifx [1]{%
 \ifx #1\expandafter \@firstoftwo
 \else \expandafter \@secondoftwo
 \fi
}%
\providecommand \natexlab [1]{#1}%
\providecommand \enquote  [1]{``#1''}%
\providecommand \bibnamefont  [1]{#1}%
\providecommand \bibfnamefont [1]{#1}%
\providecommand \citenamefont [1]{#1}%
\providecommand \href@noop [0]{\@secondoftwo}%
\providecommand \href [0]{\begingroup \@sanitize@url \@href}%
\providecommand \@href[1]{\@@startlink{#1}\@@href}%
\providecommand \@@href[1]{\endgroup#1\@@endlink}%
\providecommand \@sanitize@url [0]{\catcode `\\12\catcode `\$12\catcode
  `\&12\catcode `\#12\catcode `\^12\catcode `\_12\catcode `\%12\relax}%
\providecommand \@@startlink[1]{}%
\providecommand \@@endlink[0]{}%
\providecommand \url  [0]{\begingroup\@sanitize@url \@url }%
\providecommand \@url [1]{\endgroup\@href {#1}{\urlprefix }}%
\providecommand \urlprefix  [0]{URL }%
\providecommand \Eprint [0]{\href }%
\providecommand \doibase [0]{http://dx.doi.org/}%
\providecommand \selectlanguage [0]{\@gobble}%
\providecommand \bibinfo  [0]{\@secondoftwo}%
\providecommand \bibfield  [0]{\@secondoftwo}%
\providecommand \translation [1]{[#1]}%
\providecommand \BibitemOpen [0]{}%
\providecommand \bibitemStop [0]{}%
\providecommand \bibitemNoStop [0]{.\EOS\space}%
\providecommand \EOS [0]{\spacefactor3000\relax}%
\providecommand \BibitemShut  [1]{\csname bibitem#1\endcsname}%
\let\auto@bib@innerbib\@empty
%</preamble>
\bibitem [{\citenamefont {Pellizzari}\ \emph {et~al.}(1995)\citenamefont
  {Pellizzari}, \citenamefont {Gardiner}, \citenamefont {Cirac},\ and\
  \citenamefont {Zoller}}]{pellizzari1995decoherence}%
  \BibitemOpen
  \bibfield  {author} {\bibinfo {author} {\bibfnamefont {T.}~\bibnamefont
  {Pellizzari}}, \bibinfo {author} {\bibfnamefont {S.~A.}\ \bibnamefont
  {Gardiner}}, \bibinfo {author} {\bibfnamefont {J.~I.}\ \bibnamefont {Cirac}},
  \ and\ \bibinfo {author} {\bibfnamefont {P.}~\bibnamefont {Zoller}},\
  }\href@noop {} {\bibfield  {journal} {\bibinfo  {journal} {Physical Review
  Letters}\ }\textbf {\bibinfo {volume} {75}},\ \bibinfo {pages} {3788}
  (\bibinfo {year} {1995})}\BibitemShut {NoStop}%
\bibitem [{\citenamefont {Mabuchi}\ \emph {et~al.}(2001)\citenamefont
  {Mabuchi}, \citenamefont {Armen}, \citenamefont {Lev}, \citenamefont
  {Loncar}, \citenamefont {Vuckovic}, \citenamefont {Kimble}, \citenamefont
  {Preskill}, \citenamefont {Roukes}, \citenamefont {Scherer},\ and\
  \citenamefont {van Enk}}]{mabuchi2001quantum}%
  \BibitemOpen
  \bibfield  {author} {\bibinfo {author} {\bibfnamefont {H.}~\bibnamefont
  {Mabuchi}}, \bibinfo {author} {\bibfnamefont {M.}~\bibnamefont {Armen}},
  \bibinfo {author} {\bibfnamefont {B.}~\bibnamefont {Lev}}, \bibinfo {author}
  {\bibfnamefont {M.}~\bibnamefont {Loncar}}, \bibinfo {author} {\bibfnamefont
  {J.}~\bibnamefont {Vuckovic}}, \bibinfo {author} {\bibfnamefont {H.~J.}\
  \bibnamefont {Kimble}}, \bibinfo {author} {\bibfnamefont {J.}~\bibnamefont
  {Preskill}}, \bibinfo {author} {\bibfnamefont {M.~L.}\ \bibnamefont
  {Roukes}}, \bibinfo {author} {\bibfnamefont {A.}~\bibnamefont {Scherer}}, \
  and\ \bibinfo {author} {\bibfnamefont {S.~J.}\ \bibnamefont {van Enk}},\
  }\href@noop {} {\bibfield  {journal} {\bibinfo  {journal} {Quantum Inf.
  Comput.}\ }\textbf {\bibinfo {volume} {1}},\ \bibinfo {pages} {7} (\bibinfo
  {year} {2001})}\BibitemShut {NoStop}%
\bibitem [{\citenamefont {Kimble}(2008)}]{kimble2008quantum}%
  \BibitemOpen
  \bibfield  {author} {\bibinfo {author} {\bibfnamefont {H.~J.}\ \bibnamefont
  {Kimble}},\ }\href@noop {} {\bibfield  {journal} {\bibinfo  {journal}
  {Nature}\ }\textbf {\bibinfo {volume} {453}},\ \bibinfo {pages} {1023}
  (\bibinfo {year} {2008})}\BibitemShut {NoStop}%
\bibitem [{\citenamefont {Schine}\ \emph {et~al.}(2016)\citenamefont {Schine},
  \citenamefont {Ryou}, \citenamefont {Gromov}, \citenamefont {Sommer},\ and\
  \citenamefont {Simon}}]{schine2016synthetic}%
  \BibitemOpen
  \bibfield  {author} {\bibinfo {author} {\bibfnamefont {N.}~\bibnamefont
  {Schine}}, \bibinfo {author} {\bibfnamefont {A.}~\bibnamefont {Ryou}},
  \bibinfo {author} {\bibfnamefont {A.}~\bibnamefont {Gromov}}, \bibinfo
  {author} {\bibfnamefont {A.}~\bibnamefont {Sommer}}, \ and\ \bibinfo {author}
  {\bibfnamefont {J.}~\bibnamefont {Simon}},\ }\href@noop {} {\bibfield
  {journal} {\bibinfo  {journal} {Nature}\ }\textbf {\bibinfo {volume} {534}},\
  \bibinfo {pages} {671} (\bibinfo {year} {2016})}\BibitemShut {NoStop}%
\bibitem [{\citenamefont {Chang}\ \emph {et~al.}(2018)\citenamefont {Chang},
  \citenamefont {Douglas}, \citenamefont {Gonz{\'a}lez-Tudela}, \citenamefont
  {Hung},\ and\ \citenamefont {Kimble}}]{chang2018colloquium}%
  \BibitemOpen
  \bibfield  {author} {\bibinfo {author} {\bibfnamefont {D.}~\bibnamefont
  {Chang}}, \bibinfo {author} {\bibfnamefont {J.}~\bibnamefont {Douglas}},
  \bibinfo {author} {\bibfnamefont {A.}~\bibnamefont {Gonz{\'a}lez-Tudela}},
  \bibinfo {author} {\bibfnamefont {C.-L.}\ \bibnamefont {Hung}}, \ and\
  \bibinfo {author} {\bibfnamefont {H.}~\bibnamefont {Kimble}},\ }\href@noop {}
  {\bibfield  {journal} {\bibinfo  {journal} {Reviews of Modern Physics}\
  }\textbf {\bibinfo {volume} {90}},\ \bibinfo {pages} {031002} (\bibinfo
  {year} {2018})}\BibitemShut {NoStop}%
\bibitem [{\citenamefont {Clark}\ \emph {et~al.}(2020)\citenamefont {Clark},
  \citenamefont {Schine}, \citenamefont {Baum}, \citenamefont {Jia},\ and\
  \citenamefont {Simon}}]{clark2020observation}%
  \BibitemOpen
  \bibfield  {author} {\bibinfo {author} {\bibfnamefont {L.~W.}\ \bibnamefont
  {Clark}}, \bibinfo {author} {\bibfnamefont {N.}~\bibnamefont {Schine}},
  \bibinfo {author} {\bibfnamefont {C.}~\bibnamefont {Baum}}, \bibinfo {author}
  {\bibfnamefont {N.}~\bibnamefont {Jia}}, \ and\ \bibinfo {author}
  {\bibfnamefont {J.}~\bibnamefont {Simon}},\ }\href@noop {} {\bibfield
  {journal} {\bibinfo  {journal} {Nature}\ }\textbf {\bibinfo {volume} {582}},\
  \bibinfo {pages} {41} (\bibinfo {year} {2020})}\BibitemShut {NoStop}%
\bibitem [{\citenamefont {Mabuchi}\ and\ \citenamefont
  {Doherty}(2002)}]{mabuchi2002cavity}%
  \BibitemOpen
  \bibfield  {author} {\bibinfo {author} {\bibfnamefont {H.}~\bibnamefont
  {Mabuchi}}\ and\ \bibinfo {author} {\bibfnamefont {A.}~\bibnamefont
  {Doherty}},\ }\href@noop {} {\bibfield  {journal} {\bibinfo  {journal}
  {Science}\ }\textbf {\bibinfo {volume} {298}},\ \bibinfo {pages} {1372}
  (\bibinfo {year} {2002})}\BibitemShut {NoStop}%
\bibitem [{\citenamefont {Wilk}\ \emph {et~al.}(2010)\citenamefont {Wilk},
  \citenamefont {Ga{\"e}tan}, \citenamefont {Evellin}, \citenamefont {Wolters},
  \citenamefont {Miroshnychenko}, \citenamefont {Grangier},\ and\ \citenamefont
  {Browaeys}}]{wilk2010entanglement}%
  \BibitemOpen
  \bibfield  {author} {\bibinfo {author} {\bibfnamefont {T.}~\bibnamefont
  {Wilk}}, \bibinfo {author} {\bibfnamefont {A.}~\bibnamefont {Ga{\"e}tan}},
  \bibinfo {author} {\bibfnamefont {C.}~\bibnamefont {Evellin}}, \bibinfo
  {author} {\bibfnamefont {J.}~\bibnamefont {Wolters}}, \bibinfo {author}
  {\bibfnamefont {Y.}~\bibnamefont {Miroshnychenko}}, \bibinfo {author}
  {\bibfnamefont {P.}~\bibnamefont {Grangier}}, \ and\ \bibinfo {author}
  {\bibfnamefont {A.}~\bibnamefont {Browaeys}},\ }\href@noop {} {\bibfield
  {journal} {\bibinfo  {journal} {Physical review letters}\ }\textbf {\bibinfo
  {volume} {104}},\ \bibinfo {pages} {010502} (\bibinfo {year}
  {2010})}\BibitemShut {NoStop}%
\bibitem [{\citenamefont {Schlosser}\ \emph {et~al.}(2001)\citenamefont
  {Schlosser}, \citenamefont {Reymond}, \citenamefont {Protsenko},\ and\
  \citenamefont {Grangier}}]{schlosser2001sub}%
  \BibitemOpen
  \bibfield  {author} {\bibinfo {author} {\bibfnamefont {N.}~\bibnamefont
  {Schlosser}}, \bibinfo {author} {\bibfnamefont {G.}~\bibnamefont {Reymond}},
  \bibinfo {author} {\bibfnamefont {I.}~\bibnamefont {Protsenko}}, \ and\
  \bibinfo {author} {\bibfnamefont {P.}~\bibnamefont {Grangier}},\ }\href@noop
  {} {\bibfield  {journal} {\bibinfo  {journal} {Nature}\ }\textbf {\bibinfo
  {volume} {411}},\ \bibinfo {pages} {1024} (\bibinfo {year}
  {2001})}\BibitemShut {NoStop}%
\bibitem [{\citenamefont {Isenhower}\ \emph {et~al.}(2010)\citenamefont
  {Isenhower}, \citenamefont {Urban}, \citenamefont {Zhang}, \citenamefont
  {Gill}, \citenamefont {Henage}, \citenamefont {Johnson}, \citenamefont
  {Walker},\ and\ \citenamefont {Saffman}}]{isenhower2010demonstration}%
  \BibitemOpen
  \bibfield  {author} {\bibinfo {author} {\bibfnamefont {L.}~\bibnamefont
  {Isenhower}}, \bibinfo {author} {\bibfnamefont {E.}~\bibnamefont {Urban}},
  \bibinfo {author} {\bibfnamefont {X.}~\bibnamefont {Zhang}}, \bibinfo
  {author} {\bibfnamefont {A.}~\bibnamefont {Gill}}, \bibinfo {author}
  {\bibfnamefont {T.}~\bibnamefont {Henage}}, \bibinfo {author} {\bibfnamefont
  {T.~A.}\ \bibnamefont {Johnson}}, \bibinfo {author} {\bibfnamefont
  {T.}~\bibnamefont {Walker}}, \ and\ \bibinfo {author} {\bibfnamefont
  {M.}~\bibnamefont {Saffman}},\ }\href@noop {} {\bibfield  {journal} {\bibinfo
   {journal} {Physical review letters}\ }\textbf {\bibinfo {volume} {104}},\
  \bibinfo {pages} {010503} (\bibinfo {year} {2010})}\BibitemShut {NoStop}%
\bibitem [{\citenamefont {Lukin}(2003)}]{lukin2003colloquium}%
  \BibitemOpen
  \bibfield  {author} {\bibinfo {author} {\bibfnamefont {M.}~\bibnamefont
  {Lukin}},\ }\href@noop {} {\bibfield  {journal} {\bibinfo  {journal} {Reviews
  of Modern Physics}\ }\textbf {\bibinfo {volume} {75}},\ \bibinfo {pages}
  {457} (\bibinfo {year} {2003})}\BibitemShut {NoStop}%
\bibitem [{\citenamefont {Ebert}\ \emph {et~al.}(2015)\citenamefont {Ebert},
  \citenamefont {Kwon}, \citenamefont {Walker},\ and\ \citenamefont
  {Saffman}}]{ebert2015coherence}%
  \BibitemOpen
  \bibfield  {author} {\bibinfo {author} {\bibfnamefont {M.}~\bibnamefont
  {Ebert}}, \bibinfo {author} {\bibfnamefont {M.}~\bibnamefont {Kwon}},
  \bibinfo {author} {\bibfnamefont {T.}~\bibnamefont {Walker}}, \ and\ \bibinfo
  {author} {\bibfnamefont {M.}~\bibnamefont {Saffman}},\ }\href@noop {}
  {\bibfield  {journal} {\bibinfo  {journal} {Physical Review Letters}\
  }\textbf {\bibinfo {volume} {115}},\ \bibinfo {pages} {093601} (\bibinfo
  {year} {2015})}\BibitemShut {NoStop}%
\bibitem [{\citenamefont {Zhang}\ \emph {et~al.}(2012)\citenamefont {Zhang},
  \citenamefont {Liu}, \citenamefont {Zhou}, \citenamefont {Chuu},
  \citenamefont {Loy},\ and\ \citenamefont {Du}}]{zhang2012coherent}%
  \BibitemOpen
  \bibfield  {author} {\bibinfo {author} {\bibfnamefont {S.}~\bibnamefont
  {Zhang}}, \bibinfo {author} {\bibfnamefont {C.}~\bibnamefont {Liu}}, \bibinfo
  {author} {\bibfnamefont {S.}~\bibnamefont {Zhou}}, \bibinfo {author}
  {\bibfnamefont {C.-S.}\ \bibnamefont {Chuu}}, \bibinfo {author}
  {\bibfnamefont {M.~M.}\ \bibnamefont {Loy}}, \ and\ \bibinfo {author}
  {\bibfnamefont {S.}~\bibnamefont {Du}},\ }\href@noop {} {\bibfield  {journal}
  {\bibinfo  {journal} {Physical review letters}\ }\textbf {\bibinfo {volume}
  {109}},\ \bibinfo {pages} {263601} (\bibinfo {year} {2012})}\BibitemShut
  {NoStop}%
\bibitem [{\citenamefont {Morsch}\ and\ \citenamefont
  {Oberthaler}(2006)}]{morsch2006dynamics}%
  \BibitemOpen
  \bibfield  {author} {\bibinfo {author} {\bibfnamefont {O.}~\bibnamefont
  {Morsch}}\ and\ \bibinfo {author} {\bibfnamefont {M.}~\bibnamefont
  {Oberthaler}},\ }\href@noop {} {\bibfield  {journal} {\bibinfo  {journal}
  {Reviews of modern physics}\ }\textbf {\bibinfo {volume} {78}},\ \bibinfo
  {pages} {179} (\bibinfo {year} {2006})}\BibitemShut {NoStop}%
\bibitem [{\citenamefont {Fetter}(2009)}]{fetter2009rotating}%
  \BibitemOpen
  \bibfield  {author} {\bibinfo {author} {\bibfnamefont {A.~L.}\ \bibnamefont
  {Fetter}},\ }\href@noop {} {\bibfield  {journal} {\bibinfo  {journal}
  {Reviews of Modern Physics}\ }\textbf {\bibinfo {volume} {81}},\ \bibinfo
  {pages} {647} (\bibinfo {year} {2009})}\BibitemShut {NoStop}%
\bibitem [{\citenamefont {Kaufman}\ and\ \citenamefont
  {Ni}(2021)}]{kaufman2021quantum}%
  \BibitemOpen
  \bibfield  {author} {\bibinfo {author} {\bibfnamefont {A.~M.}\ \bibnamefont
  {Kaufman}}\ and\ \bibinfo {author} {\bibfnamefont {K.-K.}\ \bibnamefont
  {Ni}},\ }\href@noop {} {\bibfield  {journal} {\bibinfo  {journal} {Nature
  Physics}\ }\textbf {\bibinfo {volume} {17}},\ \bibinfo {pages} {1324}
  (\bibinfo {year} {2021})}\BibitemShut {NoStop}%
\bibitem [{\citenamefont {Bernien}\ \emph {et~al.}(2017)\citenamefont
  {Bernien}, \citenamefont {Schwartz}, \citenamefont {Keesling}, \citenamefont
  {Levine}, \citenamefont {Omran}, \citenamefont {Pichler}, \citenamefont
  {Choi}, \citenamefont {Zibrov}, \citenamefont {Endres}, \citenamefont
  {Greiner} \emph {et~al.}}]{bernien2017probing}%
  \BibitemOpen
  \bibfield  {author} {\bibinfo {author} {\bibfnamefont {H.}~\bibnamefont
  {Bernien}}, \bibinfo {author} {\bibfnamefont {S.}~\bibnamefont {Schwartz}},
  \bibinfo {author} {\bibfnamefont {A.}~\bibnamefont {Keesling}}, \bibinfo
  {author} {\bibfnamefont {H.}~\bibnamefont {Levine}}, \bibinfo {author}
  {\bibfnamefont {A.}~\bibnamefont {Omran}}, \bibinfo {author} {\bibfnamefont
  {H.}~\bibnamefont {Pichler}}, \bibinfo {author} {\bibfnamefont
  {S.}~\bibnamefont {Choi}}, \bibinfo {author} {\bibfnamefont {A.~S.}\
  \bibnamefont {Zibrov}}, \bibinfo {author} {\bibfnamefont {M.}~\bibnamefont
  {Endres}}, \bibinfo {author} {\bibfnamefont {M.}~\bibnamefont {Greiner}},
  \emph {et~al.},\ }\href@noop {} {\bibfield  {journal} {\bibinfo  {journal}
  {Nature}\ }\textbf {\bibinfo {volume} {551}},\ \bibinfo {pages} {579}
  (\bibinfo {year} {2017})}\BibitemShut {NoStop}%
\bibitem [{\citenamefont {Kaufman}\ \emph {et~al.}(2014)\citenamefont
  {Kaufman}, \citenamefont {Lester}, \citenamefont {Reynolds}, \citenamefont
  {Wall}, \citenamefont {Foss-Feig}, \citenamefont {Hazzard}, \citenamefont
  {Rey},\ and\ \citenamefont {Regal}}]{kaufman2014two}%
  \BibitemOpen
  \bibfield  {author} {\bibinfo {author} {\bibfnamefont {A.}~\bibnamefont
  {Kaufman}}, \bibinfo {author} {\bibfnamefont {B.}~\bibnamefont {Lester}},
  \bibinfo {author} {\bibfnamefont {C.}~\bibnamefont {Reynolds}}, \bibinfo
  {author} {\bibfnamefont {M.}~\bibnamefont {Wall}}, \bibinfo {author}
  {\bibfnamefont {M.}~\bibnamefont {Foss-Feig}}, \bibinfo {author}
  {\bibfnamefont {K.}~\bibnamefont {Hazzard}}, \bibinfo {author} {\bibfnamefont
  {A.}~\bibnamefont {Rey}}, \ and\ \bibinfo {author} {\bibfnamefont
  {C.}~\bibnamefont {Regal}},\ }\href@noop {} {\bibfield  {journal} {\bibinfo
  {journal} {Science}\ }\textbf {\bibinfo {volume} {345}},\ \bibinfo {pages}
  {306} (\bibinfo {year} {2014})}\BibitemShut {NoStop}%
\bibitem [{\citenamefont {Levine}\ \emph {et~al.}(2018)\citenamefont {Levine},
  \citenamefont {Keesling}, \citenamefont {Omran}, \citenamefont {Bernien},
  \citenamefont {Schwartz}, \citenamefont {Zibrov}, \citenamefont {Endres},
  \citenamefont {Greiner}, \citenamefont {Vuleti{\'c}},\ and\ \citenamefont
  {Lukin}}]{levine2018high}%
  \BibitemOpen
  \bibfield  {author} {\bibinfo {author} {\bibfnamefont {H.}~\bibnamefont
  {Levine}}, \bibinfo {author} {\bibfnamefont {A.}~\bibnamefont {Keesling}},
  \bibinfo {author} {\bibfnamefont {A.}~\bibnamefont {Omran}}, \bibinfo
  {author} {\bibfnamefont {H.}~\bibnamefont {Bernien}}, \bibinfo {author}
  {\bibfnamefont {S.}~\bibnamefont {Schwartz}}, \bibinfo {author}
  {\bibfnamefont {A.~S.}\ \bibnamefont {Zibrov}}, \bibinfo {author}
  {\bibfnamefont {M.}~\bibnamefont {Endres}}, \bibinfo {author} {\bibfnamefont
  {M.}~\bibnamefont {Greiner}}, \bibinfo {author} {\bibfnamefont
  {V.}~\bibnamefont {Vuleti{\'c}}}, \ and\ \bibinfo {author} {\bibfnamefont
  {M.~D.}\ \bibnamefont {Lukin}},\ }\href@noop {} {\bibfield  {journal}
  {\bibinfo  {journal} {Physical review letters}\ }\textbf {\bibinfo {volume}
  {121}},\ \bibinfo {pages} {123603} (\bibinfo {year} {2018})}\BibitemShut
  {NoStop}%
\bibitem [{\citenamefont {Xia}\ \emph {et~al.}(2015)\citenamefont {Xia},
  \citenamefont {Lichtman}, \citenamefont {Maller}, \citenamefont {Carr},
  \citenamefont {Piotrowicz}, \citenamefont {Isenhower},\ and\ \citenamefont
  {Saffman}}]{xia2015randomized}%
  \BibitemOpen
  \bibfield  {author} {\bibinfo {author} {\bibfnamefont {T.}~\bibnamefont
  {Xia}}, \bibinfo {author} {\bibfnamefont {M.}~\bibnamefont {Lichtman}},
  \bibinfo {author} {\bibfnamefont {K.}~\bibnamefont {Maller}}, \bibinfo
  {author} {\bibfnamefont {A.}~\bibnamefont {Carr}}, \bibinfo {author}
  {\bibfnamefont {M.}~\bibnamefont {Piotrowicz}}, \bibinfo {author}
  {\bibfnamefont {L.}~\bibnamefont {Isenhower}}, \ and\ \bibinfo {author}
  {\bibfnamefont {M.}~\bibnamefont {Saffman}},\ }\href@noop {} {\bibfield
  {journal} {\bibinfo  {journal} {Physical review letters}\ }\textbf {\bibinfo
  {volume} {114}},\ \bibinfo {pages} {100503} (\bibinfo {year}
  {2015})}\BibitemShut {NoStop}%
\bibitem [{\citenamefont {Duan}\ \emph {et~al.}(2000)\citenamefont {Duan},
  \citenamefont {Cirac}, \citenamefont {Zoller},\ and\ \citenamefont
  {Polzik}}]{duan2000quantum}%
  \BibitemOpen
  \bibfield  {author} {\bibinfo {author} {\bibfnamefont {L.-M.}\ \bibnamefont
  {Duan}}, \bibinfo {author} {\bibfnamefont {J.}~\bibnamefont {Cirac}},
  \bibinfo {author} {\bibfnamefont {P.}~\bibnamefont {Zoller}}, \ and\ \bibinfo
  {author} {\bibfnamefont {E.}~\bibnamefont {Polzik}},\ }\href@noop {}
  {\bibfield  {journal} {\bibinfo  {journal} {Physical review letters}\
  }\textbf {\bibinfo {volume} {85}},\ \bibinfo {pages} {5643} (\bibinfo {year}
  {2000})}\BibitemShut {NoStop}%
\bibitem [{\citenamefont {Buluta}\ and\ \citenamefont
  {Nori}(2009)}]{buluta2009quantum}%
  \BibitemOpen
  \bibfield  {author} {\bibinfo {author} {\bibfnamefont {I.}~\bibnamefont
  {Buluta}}\ and\ \bibinfo {author} {\bibfnamefont {F.}~\bibnamefont {Nori}},\
  }\href@noop {} {\bibfield  {journal} {\bibinfo  {journal} {Science}\ }\textbf
  {\bibinfo {volume} {326}},\ \bibinfo {pages} {108} (\bibinfo {year}
  {2009})}\BibitemShut {NoStop}%
\bibitem [{\citenamefont {Lewenstein}\ \emph {et~al.}(2007)\citenamefont
  {Lewenstein}, \citenamefont {Sanpera}, \citenamefont {Ahufinger},
  \citenamefont {Damski}, \citenamefont {Sen},\ and\ \citenamefont
  {Sen}}]{lewenstein2007ultracold}%
  \BibitemOpen
  \bibfield  {author} {\bibinfo {author} {\bibfnamefont {M.}~\bibnamefont
  {Lewenstein}}, \bibinfo {author} {\bibfnamefont {A.}~\bibnamefont {Sanpera}},
  \bibinfo {author} {\bibfnamefont {V.}~\bibnamefont {Ahufinger}}, \bibinfo
  {author} {\bibfnamefont {B.}~\bibnamefont {Damski}}, \bibinfo {author}
  {\bibfnamefont {A.}~\bibnamefont {Sen}}, \ and\ \bibinfo {author}
  {\bibfnamefont {U.}~\bibnamefont {Sen}},\ }\href@noop {} {\bibfield
  {journal} {\bibinfo  {journal} {Advances in Physics}\ }\textbf {\bibinfo
  {volume} {56}},\ \bibinfo {pages} {243} (\bibinfo {year} {2007})}\BibitemShut
  {NoStop}%
\bibitem [{\citenamefont {Jaksch}\ and\ \citenamefont
  {Zoller}(2005)}]{jaksch2005cold}%
  \BibitemOpen
  \bibfield  {author} {\bibinfo {author} {\bibfnamefont {D.}~\bibnamefont
  {Jaksch}}\ and\ \bibinfo {author} {\bibfnamefont {P.}~\bibnamefont
  {Zoller}},\ }\href@noop {} {\bibfield  {journal} {\bibinfo  {journal} {Annals
  of physics}\ }\textbf {\bibinfo {volume} {315}},\ \bibinfo {pages} {52}
  (\bibinfo {year} {2005})}\BibitemShut {NoStop}%
\bibitem [{\citenamefont {Kurizki}\ \emph {et~al.}(2015)\citenamefont
  {Kurizki}, \citenamefont {Bertet}, \citenamefont {Kubo}, \citenamefont
  {M{\o}lmer}, \citenamefont {Petrosyan}, \citenamefont {Rabl},\ and\
  \citenamefont {Schmiedmayer}}]{kurizki2015quantum}%
  \BibitemOpen
  \bibfield  {author} {\bibinfo {author} {\bibfnamefont {G.}~\bibnamefont
  {Kurizki}}, \bibinfo {author} {\bibfnamefont {P.}~\bibnamefont {Bertet}},
  \bibinfo {author} {\bibfnamefont {Y.}~\bibnamefont {Kubo}}, \bibinfo {author}
  {\bibfnamefont {K.}~\bibnamefont {M{\o}lmer}}, \bibinfo {author}
  {\bibfnamefont {D.}~\bibnamefont {Petrosyan}}, \bibinfo {author}
  {\bibfnamefont {P.}~\bibnamefont {Rabl}}, \ and\ \bibinfo {author}
  {\bibfnamefont {J.}~\bibnamefont {Schmiedmayer}},\ }\href@noop {} {\bibfield
  {journal} {\bibinfo  {journal} {Proceedings of the National Academy of
  Sciences}\ }\textbf {\bibinfo {volume} {112}},\ \bibinfo {pages} {3866}
  (\bibinfo {year} {2015})}\BibitemShut {NoStop}%
\bibitem [{\citenamefont {Chen}\ \emph {et~al.}(2015)\citenamefont {Chen},
  \citenamefont {Hu}, \citenamefont {Duan}, \citenamefont {Braverman},
  \citenamefont {Zhang},\ and\ \citenamefont {Vuleti{\'c}}}]{chen2015carving}%
  \BibitemOpen
  \bibfield  {author} {\bibinfo {author} {\bibfnamefont {W.}~\bibnamefont
  {Chen}}, \bibinfo {author} {\bibfnamefont {J.}~\bibnamefont {Hu}}, \bibinfo
  {author} {\bibfnamefont {Y.}~\bibnamefont {Duan}}, \bibinfo {author}
  {\bibfnamefont {B.}~\bibnamefont {Braverman}}, \bibinfo {author}
  {\bibfnamefont {H.}~\bibnamefont {Zhang}}, \ and\ \bibinfo {author}
  {\bibfnamefont {V.}~\bibnamefont {Vuleti{\'c}}},\ }\href@noop {} {\bibfield
  {journal} {\bibinfo  {journal} {Physical review letters}\ }\textbf {\bibinfo
  {volume} {115}},\ \bibinfo {pages} {250502} (\bibinfo {year}
  {2015})}\BibitemShut {NoStop}%
\bibitem [{\citenamefont {Vaneecloo}\ \emph {et~al.}(2022)\citenamefont
  {Vaneecloo}, \citenamefont {Garcia},\ and\ \citenamefont
  {Ourjoumtsev}}]{vaneecloo2022intracavity}%
  \BibitemOpen
  \bibfield  {author} {\bibinfo {author} {\bibfnamefont {J.}~\bibnamefont
  {Vaneecloo}}, \bibinfo {author} {\bibfnamefont {S.}~\bibnamefont {Garcia}}, \
  and\ \bibinfo {author} {\bibfnamefont {A.}~\bibnamefont {Ourjoumtsev}},\
  }\href@noop {} {\bibfield  {journal} {\bibinfo  {journal} {Physical Review
  X}\ }\textbf {\bibinfo {volume} {12}},\ \bibinfo {pages} {021034} (\bibinfo
  {year} {2022})}\BibitemShut {NoStop}%
\bibitem [{\citenamefont {Ramette}\ \emph {et~al.}(2022)\citenamefont
  {Ramette}, \citenamefont {Sinclair}, \citenamefont {Vendeiro}, \citenamefont
  {Rudelis}, \citenamefont {Cetina},\ and\ \citenamefont
  {Vuleti{\'c}}}]{ramette2022any}%
  \BibitemOpen
  \bibfield  {author} {\bibinfo {author} {\bibfnamefont {J.}~\bibnamefont
  {Ramette}}, \bibinfo {author} {\bibfnamefont {J.}~\bibnamefont {Sinclair}},
  \bibinfo {author} {\bibfnamefont {Z.}~\bibnamefont {Vendeiro}}, \bibinfo
  {author} {\bibfnamefont {A.}~\bibnamefont {Rudelis}}, \bibinfo {author}
  {\bibfnamefont {M.}~\bibnamefont {Cetina}}, \ and\ \bibinfo {author}
  {\bibfnamefont {V.}~\bibnamefont {Vuleti{\'c}}},\ }\href@noop {} {\bibfield
  {journal} {\bibinfo  {journal} {PRX Quantum}\ }\textbf {\bibinfo {volume}
  {3}},\ \bibinfo {pages} {010344} (\bibinfo {year} {2022})}\BibitemShut
  {NoStop}%
\bibitem [{\citenamefont {Ritter}\ \emph {et~al.}(2012)\citenamefont {Ritter},
  \citenamefont {N{\"o}lleke}, \citenamefont {Hahn}, \citenamefont {Reiserer},
  \citenamefont {Neuzner}, \citenamefont {Uphoff}, \citenamefont {M{\"u}cke},
  \citenamefont {Figueroa}, \citenamefont {Bochmann},\ and\ \citenamefont
  {Rempe}}]{ritter2012elementary}%
  \BibitemOpen
  \bibfield  {author} {\bibinfo {author} {\bibfnamefont {S.}~\bibnamefont
  {Ritter}}, \bibinfo {author} {\bibfnamefont {C.}~\bibnamefont {N{\"o}lleke}},
  \bibinfo {author} {\bibfnamefont {C.}~\bibnamefont {Hahn}}, \bibinfo {author}
  {\bibfnamefont {A.}~\bibnamefont {Reiserer}}, \bibinfo {author}
  {\bibfnamefont {A.}~\bibnamefont {Neuzner}}, \bibinfo {author} {\bibfnamefont
  {M.}~\bibnamefont {Uphoff}}, \bibinfo {author} {\bibfnamefont
  {M.}~\bibnamefont {M{\"u}cke}}, \bibinfo {author} {\bibfnamefont
  {E.}~\bibnamefont {Figueroa}}, \bibinfo {author} {\bibfnamefont
  {J.}~\bibnamefont {Bochmann}}, \ and\ \bibinfo {author} {\bibfnamefont
  {G.}~\bibnamefont {Rempe}},\ }\href@noop {} {\bibfield  {journal} {\bibinfo
  {journal} {Nature}\ }\textbf {\bibinfo {volume} {484}},\ \bibinfo {pages}
  {195} (\bibinfo {year} {2012})}\BibitemShut {NoStop}%
\bibitem [{\citenamefont {Chen}\ \emph {et~al.}(2022)\citenamefont {Chen},
  \citenamefont {Szurek}, \citenamefont {Hu}, \citenamefont {de~Hond},
  \citenamefont {Braverman},\ and\ \citenamefont {Vuletic}}]{chen2022high}%
  \BibitemOpen
  \bibfield  {author} {\bibinfo {author} {\bibfnamefont {Y.-T.}\ \bibnamefont
  {Chen}}, \bibinfo {author} {\bibfnamefont {M.}~\bibnamefont {Szurek}},
  \bibinfo {author} {\bibfnamefont {B.}~\bibnamefont {Hu}}, \bibinfo {author}
  {\bibfnamefont {J.}~\bibnamefont {de~Hond}}, \bibinfo {author} {\bibfnamefont
  {B.}~\bibnamefont {Braverman}}, \ and\ \bibinfo {author} {\bibfnamefont
  {V.}~\bibnamefont {Vuletic}},\ }\href@noop {} {\bibfield  {journal} {\bibinfo
   {journal} {arXiv preprint arXiv:2207.06876}\ } (\bibinfo {year}
  {2022})}\BibitemShut {NoStop}%
\bibitem [{\citenamefont {Utama}\ \emph {et~al.}(2021)\citenamefont {Utama},
  \citenamefont {Chow}, \citenamefont {Nguyen},\ and\ \citenamefont
  {Kurtsiefer}}]{utama2021coupling}%
  \BibitemOpen
  \bibfield  {author} {\bibinfo {author} {\bibfnamefont {A.~N.}\ \bibnamefont
  {Utama}}, \bibinfo {author} {\bibfnamefont {C.~H.}\ \bibnamefont {Chow}},
  \bibinfo {author} {\bibfnamefont {C.~H.}\ \bibnamefont {Nguyen}}, \ and\
  \bibinfo {author} {\bibfnamefont {C.}~\bibnamefont {Kurtsiefer}},\
  }\href@noop {} {\bibfield  {journal} {\bibinfo  {journal} {Optics Express}\
  }\textbf {\bibinfo {volume} {29}},\ \bibinfo {pages} {8130} (\bibinfo {year}
  {2021})}\BibitemShut {NoStop}%
\bibitem [{\citenamefont {Hunger}\ \emph {et~al.}(2010)\citenamefont {Hunger},
  \citenamefont {Steinmetz}, \citenamefont {Colombe}, \citenamefont {Deutsch},
  \citenamefont {H{\"a}nsch},\ and\ \citenamefont {Reichel}}]{hunger2010fiber}%
  \BibitemOpen
  \bibfield  {author} {\bibinfo {author} {\bibfnamefont {D.}~\bibnamefont
  {Hunger}}, \bibinfo {author} {\bibfnamefont {T.}~\bibnamefont {Steinmetz}},
  \bibinfo {author} {\bibfnamefont {Y.}~\bibnamefont {Colombe}}, \bibinfo
  {author} {\bibfnamefont {C.}~\bibnamefont {Deutsch}}, \bibinfo {author}
  {\bibfnamefont {T.~W.}\ \bibnamefont {H{\"a}nsch}}, \ and\ \bibinfo {author}
  {\bibfnamefont {J.}~\bibnamefont {Reichel}},\ }\href@noop {} {\bibfield
  {journal} {\bibinfo  {journal} {New Journal of Physics}\ }\textbf {\bibinfo
  {volume} {12}},\ \bibinfo {pages} {065038} (\bibinfo {year}
  {2010})}\BibitemShut {NoStop}%
\bibitem [{\citenamefont {Brekenfeld}\ \emph {et~al.}(2020)\citenamefont
  {Brekenfeld}, \citenamefont {Niemietz}, \citenamefont {Christesen},\ and\
  \citenamefont {Rempe}}]{brekenfeld2020quantum}%
  \BibitemOpen
  \bibfield  {author} {\bibinfo {author} {\bibfnamefont {M.}~\bibnamefont
  {Brekenfeld}}, \bibinfo {author} {\bibfnamefont {D.}~\bibnamefont
  {Niemietz}}, \bibinfo {author} {\bibfnamefont {J.~D.}\ \bibnamefont
  {Christesen}}, \ and\ \bibinfo {author} {\bibfnamefont {G.}~\bibnamefont
  {Rempe}},\ }\href@noop {} {\bibfield  {journal} {\bibinfo  {journal} {Nature
  Physics}\ }\textbf {\bibinfo {volume} {16}},\ \bibinfo {pages} {647}
  (\bibinfo {year} {2020})}\BibitemShut {NoStop}%
\bibitem [{\citenamefont {Wang}\ \emph {et~al.}(2019)\citenamefont {Wang},
  \citenamefont {Zhang}, \citenamefont {Li}, \citenamefont {Li}, \citenamefont
  {Shi},\ and\ \citenamefont {Zhang}}]{wang2019experimental}%
  \BibitemOpen
  \bibfield  {author} {\bibinfo {author} {\bibfnamefont {Y.}~\bibnamefont
  {Wang}}, \bibinfo {author} {\bibfnamefont {R.}~\bibnamefont {Zhang}},
  \bibinfo {author} {\bibfnamefont {W.}~\bibnamefont {Li}}, \bibinfo {author}
  {\bibfnamefont {B.}~\bibnamefont {Li}}, \bibinfo {author} {\bibfnamefont
  {W.}~\bibnamefont {Shi}}, \ and\ \bibinfo {author} {\bibfnamefont
  {Y.}~\bibnamefont {Zhang}},\ }\href@noop {} {\bibfield  {journal} {\bibinfo
  {journal} {Optics \& Laser Technology}\ }\textbf {\bibinfo {volume} {109}},\
  \bibinfo {pages} {348} (\bibinfo {year} {2019})}\BibitemShut {NoStop}%
\bibitem [{\citenamefont {Jaffe}\ \emph {et~al.}(2021)\citenamefont {Jaffe},
  \citenamefont {Palm}, \citenamefont {Baum}, \citenamefont {Taneja},\ and\
  \citenamefont {Simon}}]{jaffe2021aberrated}%
  \BibitemOpen
  \bibfield  {author} {\bibinfo {author} {\bibfnamefont {M.}~\bibnamefont
  {Jaffe}}, \bibinfo {author} {\bibfnamefont {L.}~\bibnamefont {Palm}},
  \bibinfo {author} {\bibfnamefont {C.}~\bibnamefont {Baum}}, \bibinfo {author}
  {\bibfnamefont {L.}~\bibnamefont {Taneja}}, \ and\ \bibinfo {author}
  {\bibfnamefont {J.}~\bibnamefont {Simon}},\ }\href@noop {} {\bibfield
  {journal} {\bibinfo  {journal} {Physical Review A}\ }\textbf {\bibinfo
  {volume} {104}},\ \bibinfo {pages} {013524} (\bibinfo {year}
  {2021})}\BibitemShut {NoStop}%
\bibitem [{\citenamefont {Stone}\ \emph {et~al.}(2021)\citenamefont {Stone},
  \citenamefont {Suleymanzade}, \citenamefont {Taneja}, \citenamefont
  {Schuster},\ and\ \citenamefont {Simon}}]{stone2021optical}%
  \BibitemOpen
  \bibfield  {author} {\bibinfo {author} {\bibfnamefont {M.}~\bibnamefont
  {Stone}}, \bibinfo {author} {\bibfnamefont {A.}~\bibnamefont {Suleymanzade}},
  \bibinfo {author} {\bibfnamefont {L.}~\bibnamefont {Taneja}}, \bibinfo
  {author} {\bibfnamefont {D.~I.}\ \bibnamefont {Schuster}}, \ and\ \bibinfo
  {author} {\bibfnamefont {J.}~\bibnamefont {Simon}},\ }\href@noop {}
  {\bibfield  {journal} {\bibinfo  {journal} {Optics Letters}\ }\textbf
  {\bibinfo {volume} {46}},\ \bibinfo {pages} {21} (\bibinfo {year}
  {2021})}\BibitemShut {NoStop}%
\bibitem [{\citenamefont {McKenna}\ \emph {et~al.}(2020)\citenamefont
  {McKenna}, \citenamefont {Witmer}, \citenamefont {Patel}, \citenamefont
  {Jiang}, \citenamefont {Van~Laer}, \citenamefont {Arrangoiz-Arriola},
  \citenamefont {Wollack}, \citenamefont {Herrmann},\ and\ \citenamefont
  {Safavi-Naeini}}]{mckenna2020cryogenic}%
  \BibitemOpen
  \bibfield  {author} {\bibinfo {author} {\bibfnamefont {T.~P.}\ \bibnamefont
  {McKenna}}, \bibinfo {author} {\bibfnamefont {J.~D.}\ \bibnamefont {Witmer}},
  \bibinfo {author} {\bibfnamefont {R.~N.}\ \bibnamefont {Patel}}, \bibinfo
  {author} {\bibfnamefont {W.}~\bibnamefont {Jiang}}, \bibinfo {author}
  {\bibfnamefont {R.}~\bibnamefont {Van~Laer}}, \bibinfo {author}
  {\bibfnamefont {P.}~\bibnamefont {Arrangoiz-Arriola}}, \bibinfo {author}
  {\bibfnamefont {E.~A.}\ \bibnamefont {Wollack}}, \bibinfo {author}
  {\bibfnamefont {J.~F.}\ \bibnamefont {Herrmann}}, \ and\ \bibinfo {author}
  {\bibfnamefont {A.~H.}\ \bibnamefont {Safavi-Naeini}},\ }\href@noop {}
  {\bibfield  {journal} {\bibinfo  {journal} {Optica}\ }\textbf {\bibinfo
  {volume} {7}},\ \bibinfo {pages} {1737} (\bibinfo {year} {2020})}\BibitemShut
  {NoStop}%
\bibitem [{\citenamefont {Kumar}\ \emph
  {et~al.}(2022{\natexlab{a}})\citenamefont {Kumar}, \citenamefont
  {Suleymanzade}, \citenamefont {Stone}, \citenamefont {Taneja}, \citenamefont
  {Anferov}, \citenamefont {Schuster},\ and\ \citenamefont
  {Simon}}]{kumar2022quantum}%
  \BibitemOpen
  \bibfield  {author} {\bibinfo {author} {\bibfnamefont {A.}~\bibnamefont
  {Kumar}}, \bibinfo {author} {\bibfnamefont {A.}~\bibnamefont {Suleymanzade}},
  \bibinfo {author} {\bibfnamefont {M.}~\bibnamefont {Stone}}, \bibinfo
  {author} {\bibfnamefont {L.}~\bibnamefont {Taneja}}, \bibinfo {author}
  {\bibfnamefont {A.}~\bibnamefont {Anferov}}, \bibinfo {author} {\bibfnamefont
  {D.~I.}\ \bibnamefont {Schuster}}, \ and\ \bibinfo {author} {\bibfnamefont
  {J.}~\bibnamefont {Simon}},\ }\href@noop {} {\bibfield  {journal} {\bibinfo
  {journal} {arXiv preprint arXiv:2207.10121}\ } (\bibinfo {year}
  {2022}{\natexlab{a}})}\BibitemShut {NoStop}%
\bibitem [{\citenamefont {Gr{\"o}blacher}\ \emph {et~al.}(2009)\citenamefont
  {Gr{\"o}blacher}, \citenamefont {Hertzberg}, \citenamefont {Vanner},
  \citenamefont {Cole}, \citenamefont {Gigan}, \citenamefont {Schwab},\ and\
  \citenamefont {Aspelmeyer}}]{groblacher2009demonstration}%
  \BibitemOpen
  \bibfield  {author} {\bibinfo {author} {\bibfnamefont {S.}~\bibnamefont
  {Gr{\"o}blacher}}, \bibinfo {author} {\bibfnamefont {J.~B.}\ \bibnamefont
  {Hertzberg}}, \bibinfo {author} {\bibfnamefont {M.~R.}\ \bibnamefont
  {Vanner}}, \bibinfo {author} {\bibfnamefont {G.~D.}\ \bibnamefont {Cole}},
  \bibinfo {author} {\bibfnamefont {S.}~\bibnamefont {Gigan}}, \bibinfo
  {author} {\bibfnamefont {K.}~\bibnamefont {Schwab}}, \ and\ \bibinfo {author}
  {\bibfnamefont {M.}~\bibnamefont {Aspelmeyer}},\ }\href@noop {} {\bibfield
  {journal} {\bibinfo  {journal} {Nature Physics}\ }\textbf {\bibinfo {volume}
  {5}},\ \bibinfo {pages} {485} (\bibinfo {year} {2009})}\BibitemShut {NoStop}%
\bibitem [{\citenamefont {Falke}\ \emph {et~al.}(2014)\citenamefont {Falke},
  \citenamefont {Lemke}, \citenamefont {Grebing}, \citenamefont {Lipphardt},
  \citenamefont {Weyers}, \citenamefont {Gerginov}, \citenamefont {Huntemann},
  \citenamefont {Hagemann}, \citenamefont {Al-Masoudi}, \citenamefont
  {H{\"a}fner} \emph {et~al.}}]{falke2014strontium}%
  \BibitemOpen
  \bibfield  {author} {\bibinfo {author} {\bibfnamefont {S.}~\bibnamefont
  {Falke}}, \bibinfo {author} {\bibfnamefont {N.}~\bibnamefont {Lemke}},
  \bibinfo {author} {\bibfnamefont {C.}~\bibnamefont {Grebing}}, \bibinfo
  {author} {\bibfnamefont {B.}~\bibnamefont {Lipphardt}}, \bibinfo {author}
  {\bibfnamefont {S.}~\bibnamefont {Weyers}}, \bibinfo {author} {\bibfnamefont
  {V.}~\bibnamefont {Gerginov}}, \bibinfo {author} {\bibfnamefont
  {N.}~\bibnamefont {Huntemann}}, \bibinfo {author} {\bibfnamefont
  {C.}~\bibnamefont {Hagemann}}, \bibinfo {author} {\bibfnamefont
  {A.}~\bibnamefont {Al-Masoudi}}, \bibinfo {author} {\bibfnamefont
  {S.}~\bibnamefont {H{\"a}fner}},  \emph {et~al.},\ }\href@noop {} {\bibfield
  {journal} {\bibinfo  {journal} {New Journal of Physics}\ }\textbf {\bibinfo
  {volume} {16}},\ \bibinfo {pages} {073023} (\bibinfo {year}
  {2014})}\BibitemShut {NoStop}%
\bibitem [{\citenamefont {Yang}\ \emph {et~al.}(2016)\citenamefont {Yang},
  \citenamefont {Wang}, \citenamefont {Bao},\ and\ \citenamefont
  {Pan}}]{yang2016efficient}%
  \BibitemOpen
  \bibfield  {author} {\bibinfo {author} {\bibfnamefont {S.-J.}\ \bibnamefont
  {Yang}}, \bibinfo {author} {\bibfnamefont {X.-J.}\ \bibnamefont {Wang}},
  \bibinfo {author} {\bibfnamefont {X.-H.}\ \bibnamefont {Bao}}, \ and\
  \bibinfo {author} {\bibfnamefont {J.-W.}\ \bibnamefont {Pan}},\ }\href@noop
  {} {\bibfield  {journal} {\bibinfo  {journal} {Nature Photonics}\ }\textbf
  {\bibinfo {volume} {10}},\ \bibinfo {pages} {381} (\bibinfo {year}
  {2016})}\BibitemShut {NoStop}%
\bibitem [{\citenamefont {Yang}\ \emph {et~al.}(2021)\citenamefont {Yang},
  \citenamefont {Li}, \citenamefont {Zhou}, \citenamefont {Jiang},
  \citenamefont {Bao},\ and\ \citenamefont {Pan}}]{yang2021single}%
  \BibitemOpen
  \bibfield  {author} {\bibinfo {author} {\bibfnamefont {C.-W.}\ \bibnamefont
  {Yang}}, \bibinfo {author} {\bibfnamefont {J.}~\bibnamefont {Li}}, \bibinfo
  {author} {\bibfnamefont {M.-T.}\ \bibnamefont {Zhou}}, \bibinfo {author}
  {\bibfnamefont {X.}~\bibnamefont {Jiang}}, \bibinfo {author} {\bibfnamefont
  {X.-H.}\ \bibnamefont {Bao}}, \ and\ \bibinfo {author} {\bibfnamefont
  {J.-W.}\ \bibnamefont {Pan}},\ }\href@noop {} {\bibfield  {journal} {\bibinfo
   {journal} {arXiv preprint arXiv:2106.10858}\ } (\bibinfo {year}
  {2021})}\BibitemShut {NoStop}%
\bibitem [{\citenamefont {Kirby}\ \emph {et~al.}(1993)\citenamefont {Kirby},
  \citenamefont {Collet},\ and\ \citenamefont {Skarpaas}}]{kirby1993situ}%
  \BibitemOpen
  \bibfield  {author} {\bibinfo {author} {\bibfnamefont {R.}~\bibnamefont
  {Kirby}}, \bibinfo {author} {\bibfnamefont {G.}~\bibnamefont {Collet}}, \
  and\ \bibinfo {author} {\bibfnamefont {K.}~\bibnamefont {Skarpaas}},\ }in\
  \href@noop {} {\emph {\bibinfo {booktitle} {Proceedings of International
  Conference on Particle Accelerators}}}\ (\bibinfo {organization} {IEEE},\
  \bibinfo {year} {1993})\ pp.\ \bibinfo {pages} {3030--3032}\BibitemShut
  {NoStop}%
\bibitem [{\citenamefont {Springate}\ \emph {et~al.}(2008)\citenamefont
  {Springate}, \citenamefont {Froud}, \citenamefont {Turcu}, \citenamefont
  {Spurdle}, \citenamefont {Wolff}, \citenamefont {Hook}, \citenamefont
  {Landowski}, \citenamefont {Underwood}, \citenamefont {Cavalleri},
  \citenamefont {Dhesi} \emph {et~al.}}]{springate2008artemis}%
  \BibitemOpen
  \bibfield  {author} {\bibinfo {author} {\bibfnamefont {E.}~\bibnamefont
  {Springate}}, \bibinfo {author} {\bibfnamefont {C.}~\bibnamefont {Froud}},
  \bibinfo {author} {\bibfnamefont {I.}~\bibnamefont {Turcu}}, \bibinfo
  {author} {\bibfnamefont {S.}~\bibnamefont {Spurdle}}, \bibinfo {author}
  {\bibfnamefont {D.}~\bibnamefont {Wolff}}, \bibinfo {author} {\bibfnamefont
  {S.}~\bibnamefont {Hook}}, \bibinfo {author} {\bibfnamefont {B.}~\bibnamefont
  {Landowski}}, \bibinfo {author} {\bibfnamefont {J.}~\bibnamefont
  {Underwood}}, \bibinfo {author} {\bibfnamefont {A.}~\bibnamefont
  {Cavalleri}}, \bibinfo {author} {\bibfnamefont {S.}~\bibnamefont {Dhesi}},
  \emph {et~al.},\ }\href@noop {} {\bibfield  {journal} {\bibinfo  {journal}
  {CLF Ann Report}\ }\textbf {\bibinfo {volume} {9}},\ \bibinfo {pages} {221}
  (\bibinfo {year} {2008})}\BibitemShut {NoStop}%
\bibitem [{\citenamefont {Mestres}\ \emph {et~al.}(2015)\citenamefont
  {Mestres}, \citenamefont {Berthelot}, \citenamefont {Spasenovi{\'c}},
  \citenamefont {Gieseler}, \citenamefont {Novotny},\ and\ \citenamefont
  {Quidant}}]{mestres2015long}%
  \BibitemOpen
  \bibfield  {author} {\bibinfo {author} {\bibfnamefont {P.}~\bibnamefont
  {Mestres}}, \bibinfo {author} {\bibfnamefont {J.}~\bibnamefont {Berthelot}},
  \bibinfo {author} {\bibfnamefont {M.}~\bibnamefont {Spasenovi{\'c}}},
  \bibinfo {author} {\bibfnamefont {J.}~\bibnamefont {Gieseler}}, \bibinfo
  {author} {\bibfnamefont {L.}~\bibnamefont {Novotny}}, \ and\ \bibinfo
  {author} {\bibfnamefont {R.}~\bibnamefont {Quidant}},\ }\href@noop {}
  {\bibfield  {journal} {\bibinfo  {journal} {arXiv preprint arXiv:1505.02012}\
  } (\bibinfo {year} {2015})}\BibitemShut {NoStop}%
\bibitem [{\citenamefont {Zhong}\ \emph {et~al.}(2017)\citenamefont {Zhong},
  \citenamefont {Fl{\"a}schner}, \citenamefont {Schwarz}, \citenamefont
  {Wiesendanger}, \citenamefont {Christoph}, \citenamefont {Wagner},
  \citenamefont {Bick}, \citenamefont {Staarmann}, \citenamefont {Abeln},
  \citenamefont {Sengstock} \emph {et~al.}}]{zhong2017millikelvin}%
  \BibitemOpen
  \bibfield  {author} {\bibinfo {author} {\bibfnamefont {H.}~\bibnamefont
  {Zhong}}, \bibinfo {author} {\bibfnamefont {G.}~\bibnamefont
  {Fl{\"a}schner}}, \bibinfo {author} {\bibfnamefont {A.}~\bibnamefont
  {Schwarz}}, \bibinfo {author} {\bibfnamefont {R.}~\bibnamefont
  {Wiesendanger}}, \bibinfo {author} {\bibfnamefont {P.}~\bibnamefont
  {Christoph}}, \bibinfo {author} {\bibfnamefont {T.}~\bibnamefont {Wagner}},
  \bibinfo {author} {\bibfnamefont {A.}~\bibnamefont {Bick}}, \bibinfo {author}
  {\bibfnamefont {C.}~\bibnamefont {Staarmann}}, \bibinfo {author}
  {\bibfnamefont {B.}~\bibnamefont {Abeln}}, \bibinfo {author} {\bibfnamefont
  {K.}~\bibnamefont {Sengstock}},  \emph {et~al.},\ }\href@noop {} {\bibfield
  {journal} {\bibinfo  {journal} {Review of Scientific Instruments}\ }\textbf
  {\bibinfo {volume} {88}},\ \bibinfo {pages} {023115} (\bibinfo {year}
  {2017})}\BibitemShut {NoStop}%
\bibitem [{\citenamefont {Tanji-Suzuki}\ \emph {et~al.}(2011)\citenamefont
  {Tanji-Suzuki}, \citenamefont {Leroux}, \citenamefont {Schleier-Smith},
  \citenamefont {Cetina}, \citenamefont {Grier}, \citenamefont {Simon},\ and\
  \citenamefont {Vuleti{\'c}}}]{tanji2011interaction}%
  \BibitemOpen
  \bibfield  {author} {\bibinfo {author} {\bibfnamefont {H.}~\bibnamefont
  {Tanji-Suzuki}}, \bibinfo {author} {\bibfnamefont {I.~D.}\ \bibnamefont
  {Leroux}}, \bibinfo {author} {\bibfnamefont {M.~H.}\ \bibnamefont
  {Schleier-Smith}}, \bibinfo {author} {\bibfnamefont {M.}~\bibnamefont
  {Cetina}}, \bibinfo {author} {\bibfnamefont {A.~T.}\ \bibnamefont {Grier}},
  \bibinfo {author} {\bibfnamefont {J.}~\bibnamefont {Simon}}, \ and\ \bibinfo
  {author} {\bibfnamefont {V.}~\bibnamefont {Vuleti{\'c}}},\ }in\ \href@noop {}
  {\emph {\bibinfo {booktitle} {Advances in atomic, molecular, and optical
  physics}}},\ Vol.~\bibinfo {volume} {60}\ (\bibinfo  {publisher} {Elsevier},\
  \bibinfo {year} {2011})\ pp.\ \bibinfo {pages} {201--237}\BibitemShut
  {NoStop}%
\bibitem [{\citenamefont {Hamm}\ \emph {et~al.}(1994)\citenamefont {Hamm},
  \citenamefont {Ritter},\ and\ \citenamefont {Temkin}}]{hamm1994compact}%
  \BibitemOpen
  \bibfield  {author} {\bibinfo {author} {\bibfnamefont {R.}~\bibnamefont
  {Hamm}}, \bibinfo {author} {\bibfnamefont {D.}~\bibnamefont {Ritter}}, \ and\
  \bibinfo {author} {\bibfnamefont {H.}~\bibnamefont {Temkin}},\ }\href@noop {}
  {\bibfield  {journal} {\bibinfo  {journal} {Journal of Vacuum Science \&
  Technology A: Vacuum, Surfaces, and Films}\ }\textbf {\bibinfo {volume}
  {12}},\ \bibinfo {pages} {2790} (\bibinfo {year} {1994})}\BibitemShut
  {NoStop}%
\bibitem [{\citenamefont {Celinski}(2001)}]{celinski2001molecular}%
  \BibitemOpen
  \bibfield  {author} {\bibinfo {author} {\bibfnamefont {Z.}~\bibnamefont
  {Celinski}},\ }\href@noop {} {\bibfield  {journal} {\bibinfo  {journal}
  {Journal of Vacuum Science \& Technology A: Vacuum, Surfaces, and Films}\
  }\textbf {\bibinfo {volume} {19}},\ \bibinfo {pages} {383} (\bibinfo {year}
  {2001})}\BibitemShut {NoStop}%
\bibitem [{\citenamefont {Zhang}\ \emph {et~al.}(2011)\citenamefont {Zhang},
  \citenamefont {Zhang}, \citenamefont {Ru}, \citenamefont {Chen},\ and\
  \citenamefont {Sun}}]{zhang2011load}%
  \BibitemOpen
  \bibfield  {author} {\bibinfo {author} {\bibfnamefont {Y.~L.}\ \bibnamefont
  {Zhang}}, \bibinfo {author} {\bibfnamefont {Y.}~\bibnamefont {Zhang}},
  \bibinfo {author} {\bibfnamefont {C.}~\bibnamefont {Ru}}, \bibinfo {author}
  {\bibfnamefont {B.~K.}\ \bibnamefont {Chen}}, \ and\ \bibinfo {author}
  {\bibfnamefont {Y.}~\bibnamefont {Sun}},\ }\href@noop {} {\bibfield
  {journal} {\bibinfo  {journal} {IEEE/ASME Transactions on Mechatronics}\
  }\textbf {\bibinfo {volume} {18}},\ \bibinfo {pages} {230} (\bibinfo {year}
  {2011})}\BibitemShut {NoStop}%
\bibitem [{\citenamefont {L{\"o}w}\ \emph {et~al.}(2007)\citenamefont
  {L{\"o}w}, \citenamefont {Raitzsch}, \citenamefont {Heidemann}, \citenamefont
  {Bendkowsky}, \citenamefont {Butscher}, \citenamefont {Grabowski},\ and\
  \citenamefont {Pfau}}]{low2007apparatus}%
  \BibitemOpen
  \bibfield  {author} {\bibinfo {author} {\bibfnamefont {R.}~\bibnamefont
  {L{\"o}w}}, \bibinfo {author} {\bibfnamefont {U.}~\bibnamefont {Raitzsch}},
  \bibinfo {author} {\bibfnamefont {R.}~\bibnamefont {Heidemann}}, \bibinfo
  {author} {\bibfnamefont {V.}~\bibnamefont {Bendkowsky}}, \bibinfo {author}
  {\bibfnamefont {B.}~\bibnamefont {Butscher}}, \bibinfo {author}
  {\bibfnamefont {A.}~\bibnamefont {Grabowski}}, \ and\ \bibinfo {author}
  {\bibfnamefont {T.}~\bibnamefont {Pfau}},\ }\href@noop {} {\bibfield
  {journal} {\bibinfo  {journal} {arXiv preprint arXiv:0706.2639}\ } (\bibinfo
  {year} {2007})}\BibitemShut {NoStop}%
\bibitem [{\citenamefont {L{\'e}onard}(2017)}]{leonard2017supersolid}%
  \BibitemOpen
  \bibfield  {author} {\bibinfo {author} {\bibfnamefont {J.}~\bibnamefont
  {L{\'e}onard}},\ }\emph {\bibinfo {title} {A supersolid of matter and
  light}},\ \href@noop {} {Ph.D. thesis},\ \bibinfo  {school} {ETH Zurich}
  (\bibinfo {year} {2017})\BibitemShut {NoStop}%
\bibitem [{\citenamefont {Gehm}\ \emph {et~al.}(1998)\citenamefont {Gehm},
  \citenamefont {O’hara}, \citenamefont {Savard},\ and\ \citenamefont
  {Thomas}}]{gehm1998dynamics}%
  \BibitemOpen
  \bibfield  {author} {\bibinfo {author} {\bibfnamefont {M.}~\bibnamefont
  {Gehm}}, \bibinfo {author} {\bibfnamefont {K.}~\bibnamefont {O’hara}},
  \bibinfo {author} {\bibfnamefont {T.}~\bibnamefont {Savard}}, \ and\ \bibinfo
  {author} {\bibfnamefont {J.}~\bibnamefont {Thomas}},\ }\href@noop {}
  {\bibfield  {journal} {\bibinfo  {journal} {Physical Review A}\ }\textbf
  {\bibinfo {volume} {58}},\ \bibinfo {pages} {3914} (\bibinfo {year}
  {1998})}\BibitemShut {NoStop}%
\bibitem [{Note1()}]{Note1}%
  \BibitemOpen
  \bibinfo {note} {The spring washers are important because we have, in the
  past, observed that the time-varying thermal load associated with turning the
  dispensers on and off can eventually loosen the screws and break electrical
  contact}\BibitemShut {NoStop}%
\bibitem [{\citenamefont {Jaffe}\ \emph {et~al.}(2022)\citenamefont {Jaffe},
  \citenamefont {Palm}, \citenamefont {Baum}, \citenamefont {Taneja},
  \citenamefont {Kumar},\ and\ \citenamefont {Simon}}]{jaffe2022understanding}%
  \BibitemOpen
  \bibfield  {author} {\bibinfo {author} {\bibfnamefont {M.}~\bibnamefont
  {Jaffe}}, \bibinfo {author} {\bibfnamefont {L.}~\bibnamefont {Palm}},
  \bibinfo {author} {\bibfnamefont {C.}~\bibnamefont {Baum}}, \bibinfo {author}
  {\bibfnamefont {L.}~\bibnamefont {Taneja}}, \bibinfo {author} {\bibfnamefont
  {A.}~\bibnamefont {Kumar}}, \ and\ \bibinfo {author} {\bibfnamefont
  {J.}~\bibnamefont {Simon}},\ }\href {\doibase 10.1364/OPTICA.463723}
  {\bibfield  {journal} {\bibinfo  {journal} {Optica}\ }\textbf {\bibinfo
  {volume} {9}},\ \bibinfo {pages} {878} (\bibinfo {year} {2022})}\BibitemShut
  {NoStop}%
\bibitem [{\citenamefont {Georgakopoulos}\ \emph {et~al.}(2018)\citenamefont
  {Georgakopoulos}, \citenamefont {Sommer},\ and\ \citenamefont
  {Simon}}]{georgakopoulos2018theory}%
  \BibitemOpen
  \bibfield  {author} {\bibinfo {author} {\bibfnamefont {A.}~\bibnamefont
  {Georgakopoulos}}, \bibinfo {author} {\bibfnamefont {A.}~\bibnamefont
  {Sommer}}, \ and\ \bibinfo {author} {\bibfnamefont {J.}~\bibnamefont
  {Simon}},\ }\href@noop {} {\bibfield  {journal} {\bibinfo  {journal} {Quantum
  Science and Technology}\ }\textbf {\bibinfo {volume} {4}},\ \bibinfo {pages}
  {014005} (\bibinfo {year} {2018})}\BibitemShut {NoStop}%
\bibitem [{\citenamefont {G{\"a}nger}\ \emph {et~al.}(2018)\citenamefont
  {G{\"a}nger}, \citenamefont {Phieler}, \citenamefont {Nagler},\ and\
  \citenamefont {Widera}}]{ganger2018versatile}%
  \BibitemOpen
  \bibfield  {author} {\bibinfo {author} {\bibfnamefont {B.}~\bibnamefont
  {G{\"a}nger}}, \bibinfo {author} {\bibfnamefont {J.}~\bibnamefont {Phieler}},
  \bibinfo {author} {\bibfnamefont {B.}~\bibnamefont {Nagler}}, \ and\ \bibinfo
  {author} {\bibfnamefont {A.}~\bibnamefont {Widera}},\ }\href@noop {}
  {\bibfield  {journal} {\bibinfo  {journal} {Review of Scientific
  Instruments}\ }\textbf {\bibinfo {volume} {89}},\ \bibinfo {pages} {093105}
  (\bibinfo {year} {2018})}\BibitemShut {NoStop}%
\bibitem [{\citenamefont {{\"O}ttl}\ \emph {et~al.}(2006)\citenamefont
  {{\"O}ttl}, \citenamefont {Ritter}, \citenamefont {K{\"o}hl},\ and\
  \citenamefont {Esslinger}}]{ottl2006hybrid}%
  \BibitemOpen
  \bibfield  {author} {\bibinfo {author} {\bibfnamefont {A.}~\bibnamefont
  {{\"O}ttl}}, \bibinfo {author} {\bibfnamefont {S.}~\bibnamefont {Ritter}},
  \bibinfo {author} {\bibfnamefont {M.}~\bibnamefont {K{\"o}hl}}, \ and\
  \bibinfo {author} {\bibfnamefont {T.}~\bibnamefont {Esslinger}},\ }\href@noop
  {} {\bibfield  {journal} {\bibinfo  {journal} {Review of Scientific
  Instruments}\ }\textbf {\bibinfo {volume} {77}},\ \bibinfo {pages} {063118}
  (\bibinfo {year} {2006})}\BibitemShut {NoStop}%
\bibitem [{\citenamefont {Smith}(2016)}]{smith2016editorial}%
  \BibitemOpen
  \bibfield  {author} {\bibinfo {author} {\bibfnamefont {M.}~\bibnamefont
  {Smith}},\ }\href@noop {} {\bibfield  {journal} {\bibinfo  {journal} {Phys.
  Rev. Lett}\ }\textbf {\bibinfo {volume} {117}},\ \bibinfo {pages} {100001}
  (\bibinfo {year} {2016})}\BibitemShut {NoStop}%
\bibitem [{\citenamefont {Imamo{\u{g}}lu}(2009)}]{imamouglu2009cavity}%
  \BibitemOpen
  \bibfield  {author} {\bibinfo {author} {\bibfnamefont {A.}~\bibnamefont
  {Imamo{\u{g}}lu}},\ }\href@noop {} {\bibfield  {journal} {\bibinfo  {journal}
  {Physical review letters}\ }\textbf {\bibinfo {volume} {102}},\ \bibinfo
  {pages} {083602} (\bibinfo {year} {2009})}\BibitemShut {NoStop}%
\bibitem [{\citenamefont {Raimond}\ \emph {et~al.}(2001)\citenamefont
  {Raimond}, \citenamefont {Brune},\ and\ \citenamefont
  {Haroche}}]{raimond2001manipulating}%
  \BibitemOpen
  \bibfield  {author} {\bibinfo {author} {\bibfnamefont {J.-M.}\ \bibnamefont
  {Raimond}}, \bibinfo {author} {\bibfnamefont {M.}~\bibnamefont {Brune}}, \
  and\ \bibinfo {author} {\bibfnamefont {S.}~\bibnamefont {Haroche}},\
  }\href@noop {} {\bibfield  {journal} {\bibinfo  {journal} {Reviews of Modern
  Physics}\ }\textbf {\bibinfo {volume} {73}},\ \bibinfo {pages} {565}
  (\bibinfo {year} {2001})}\BibitemShut {NoStop}%
\bibitem [{\citenamefont {Murch}\ \emph {et~al.}(2008)\citenamefont {Murch},
  \citenamefont {Moore}, \citenamefont {Gupta},\ and\ \citenamefont
  {Stamper-Kurn}}]{murch2008observation}%
  \BibitemOpen
  \bibfield  {author} {\bibinfo {author} {\bibfnamefont {K.~W.}\ \bibnamefont
  {Murch}}, \bibinfo {author} {\bibfnamefont {K.~L.}\ \bibnamefont {Moore}},
  \bibinfo {author} {\bibfnamefont {S.}~\bibnamefont {Gupta}}, \ and\ \bibinfo
  {author} {\bibfnamefont {D.~M.}\ \bibnamefont {Stamper-Kurn}},\ }\href@noop
  {} {\bibfield  {journal} {\bibinfo  {journal} {Nature Physics}\ }\textbf
  {\bibinfo {volume} {4}},\ \bibinfo {pages} {561} (\bibinfo {year}
  {2008})}\BibitemShut {NoStop}%
\bibitem [{\citenamefont {Stamper-Kurn}(2014)}]{stamper2014cavity}%
  \BibitemOpen
  \bibfield  {author} {\bibinfo {author} {\bibfnamefont {D.~M.}\ \bibnamefont
  {Stamper-Kurn}},\ }in\ \href@noop {} {\emph {\bibinfo {booktitle} {Cavity
  optomechanics}}}\ (\bibinfo  {publisher} {Springer},\ \bibinfo {year}
  {2014})\ pp.\ \bibinfo {pages} {283--325}\BibitemShut {NoStop}%
\bibitem [{\citenamefont {Brennecke}\ \emph {et~al.}(2008)\citenamefont
  {Brennecke}, \citenamefont {Ritter}, \citenamefont {Donner},\ and\
  \citenamefont {Esslinger}}]{brennecke2008cavity}%
  \BibitemOpen
  \bibfield  {author} {\bibinfo {author} {\bibfnamefont {F.}~\bibnamefont
  {Brennecke}}, \bibinfo {author} {\bibfnamefont {S.}~\bibnamefont {Ritter}},
  \bibinfo {author} {\bibfnamefont {T.}~\bibnamefont {Donner}}, \ and\ \bibinfo
  {author} {\bibfnamefont {T.}~\bibnamefont {Esslinger}},\ }\href@noop {}
  {\bibfield  {journal} {\bibinfo  {journal} {Science}\ }\textbf {\bibinfo
  {volume} {322}},\ \bibinfo {pages} {235} (\bibinfo {year}
  {2008})}\BibitemShut {NoStop}%
\bibitem [{\citenamefont {Schleier-Smith}\ \emph {et~al.}(2011)\citenamefont
  {Schleier-Smith}, \citenamefont {Leroux}, \citenamefont {Zhang},
  \citenamefont {Van~Camp},\ and\ \citenamefont
  {Vuleti{\'c}}}]{schleier2011optomechanical}%
  \BibitemOpen
  \bibfield  {author} {\bibinfo {author} {\bibfnamefont {M.~H.}\ \bibnamefont
  {Schleier-Smith}}, \bibinfo {author} {\bibfnamefont {I.~D.}\ \bibnamefont
  {Leroux}}, \bibinfo {author} {\bibfnamefont {H.}~\bibnamefont {Zhang}},
  \bibinfo {author} {\bibfnamefont {M.~A.}\ \bibnamefont {Van~Camp}}, \ and\
  \bibinfo {author} {\bibfnamefont {V.}~\bibnamefont {Vuleti{\'c}}},\
  }\href@noop {} {\bibfield  {journal} {\bibinfo  {journal} {Physical review
  letters}\ }\textbf {\bibinfo {volume} {107}},\ \bibinfo {pages} {143005}
  (\bibinfo {year} {2011})}\BibitemShut {NoStop}%
\bibitem [{\citenamefont {Thompson}\ \emph {et~al.}(2013)\citenamefont
  {Thompson}, \citenamefont {Tiecke}, \citenamefont {de~Leon}, \citenamefont
  {Feist}, \citenamefont {Akimov}, \citenamefont {Gullans}, \citenamefont
  {Zibrov}, \citenamefont {Vuleti{\'c}},\ and\ \citenamefont
  {Lukin}}]{thompson2013coupling}%
  \BibitemOpen
  \bibfield  {author} {\bibinfo {author} {\bibfnamefont {J.~D.}\ \bibnamefont
  {Thompson}}, \bibinfo {author} {\bibfnamefont {T.}~\bibnamefont {Tiecke}},
  \bibinfo {author} {\bibfnamefont {N.~P.}\ \bibnamefont {de~Leon}}, \bibinfo
  {author} {\bibfnamefont {J.}~\bibnamefont {Feist}}, \bibinfo {author}
  {\bibfnamefont {A.}~\bibnamefont {Akimov}}, \bibinfo {author} {\bibfnamefont
  {M.}~\bibnamefont {Gullans}}, \bibinfo {author} {\bibfnamefont {A.~S.}\
  \bibnamefont {Zibrov}}, \bibinfo {author} {\bibfnamefont {V.}~\bibnamefont
  {Vuleti{\'c}}}, \ and\ \bibinfo {author} {\bibfnamefont {M.~D.}\ \bibnamefont
  {Lukin}},\ }\href@noop {} {\bibfield  {journal} {\bibinfo  {journal}
  {Science}\ }\textbf {\bibinfo {volume} {340}},\ \bibinfo {pages} {1202}
  (\bibinfo {year} {2013})}\BibitemShut {NoStop}%
\bibitem [{\citenamefont {Periwal}\ \emph {et~al.}(2021)\citenamefont
  {Periwal}, \citenamefont {Cooper}, \citenamefont {Kunkel}, \citenamefont
  {Wienand}, \citenamefont {Davis},\ and\ \citenamefont
  {Schleier-Smith}}]{periwal2021programmable}%
  \BibitemOpen
  \bibfield  {author} {\bibinfo {author} {\bibfnamefont {A.}~\bibnamefont
  {Periwal}}, \bibinfo {author} {\bibfnamefont {E.~S.}\ \bibnamefont {Cooper}},
  \bibinfo {author} {\bibfnamefont {P.}~\bibnamefont {Kunkel}}, \bibinfo
  {author} {\bibfnamefont {J.~F.}\ \bibnamefont {Wienand}}, \bibinfo {author}
  {\bibfnamefont {E.~J.}\ \bibnamefont {Davis}}, \ and\ \bibinfo {author}
  {\bibfnamefont {M.}~\bibnamefont {Schleier-Smith}},\ }\href@noop {}
  {\bibfield  {journal} {\bibinfo  {journal} {Nature}\ }\textbf {\bibinfo
  {volume} {600}},\ \bibinfo {pages} {630} (\bibinfo {year}
  {2021})}\BibitemShut {NoStop}%
\bibitem [{\citenamefont {Heinz}\ \emph {et~al.}(2021)\citenamefont {Heinz},
  \citenamefont {Trautmann}, \citenamefont {{\v{S}}anti{\'c}}, \citenamefont
  {Park}, \citenamefont {Bloch},\ and\ \citenamefont
  {Blatt}}]{heinz2021crossed}%
  \BibitemOpen
  \bibfield  {author} {\bibinfo {author} {\bibfnamefont {A.}~\bibnamefont
  {Heinz}}, \bibinfo {author} {\bibfnamefont {J.}~\bibnamefont {Trautmann}},
  \bibinfo {author} {\bibfnamefont {N.}~\bibnamefont {{\v{S}}anti{\'c}}},
  \bibinfo {author} {\bibfnamefont {A.~J.}\ \bibnamefont {Park}}, \bibinfo
  {author} {\bibfnamefont {I.}~\bibnamefont {Bloch}}, \ and\ \bibinfo {author}
  {\bibfnamefont {S.}~\bibnamefont {Blatt}},\ }\href@noop {} {\bibfield
  {journal} {\bibinfo  {journal} {Optics Letters}\ }\textbf {\bibinfo {volume}
  {46}},\ \bibinfo {pages} {250} (\bibinfo {year} {2021})}\BibitemShut
  {NoStop}%
\bibitem [{\citenamefont {Nogues}\ \emph {et~al.}(1999)\citenamefont {Nogues},
  \citenamefont {Rauschenbeutel}, \citenamefont {Osnaghi}, \citenamefont
  {Brune}, \citenamefont {Raimond},\ and\ \citenamefont
  {Haroche}}]{nogues1999seeing}%
  \BibitemOpen
  \bibfield  {author} {\bibinfo {author} {\bibfnamefont {G.}~\bibnamefont
  {Nogues}}, \bibinfo {author} {\bibfnamefont {A.}~\bibnamefont
  {Rauschenbeutel}}, \bibinfo {author} {\bibfnamefont {S.}~\bibnamefont
  {Osnaghi}}, \bibinfo {author} {\bibfnamefont {M.}~\bibnamefont {Brune}},
  \bibinfo {author} {\bibfnamefont {J.-M.}\ \bibnamefont {Raimond}}, \ and\
  \bibinfo {author} {\bibfnamefont {S.}~\bibnamefont {Haroche}},\ }\href@noop
  {} {\bibfield  {journal} {\bibinfo  {journal} {Nature}\ }\textbf {\bibinfo
  {volume} {400}},\ \bibinfo {pages} {239} (\bibinfo {year}
  {1999})}\BibitemShut {NoStop}%
\bibitem [{\citenamefont {Turchette}\ \emph {et~al.}(1995)\citenamefont
  {Turchette}, \citenamefont {Thompson},\ and\ \citenamefont
  {Kimble}}]{turchette1995one}%
  \BibitemOpen
  \bibfield  {author} {\bibinfo {author} {\bibfnamefont {Q.}~\bibnamefont
  {Turchette}}, \bibinfo {author} {\bibfnamefont {R.}~\bibnamefont {Thompson}},
  \ and\ \bibinfo {author} {\bibfnamefont {H.}~\bibnamefont {Kimble}},\
  }\href@noop {} {\bibfield  {journal} {\bibinfo  {journal} {Applied physics
  B-lasers and optics}\ }\textbf {\bibinfo {volume} {60}},\ \bibinfo {pages}
  {S1} (\bibinfo {year} {1995})}\BibitemShut {NoStop}%
\bibitem [{\citenamefont {Mabuchi}\ \emph {et~al.}(1996)\citenamefont
  {Mabuchi}, \citenamefont {Turchette}, \citenamefont {Chapman},\ and\
  \citenamefont {Kimble}}]{mabuchi1996real}%
  \BibitemOpen
  \bibfield  {author} {\bibinfo {author} {\bibfnamefont {H.}~\bibnamefont
  {Mabuchi}}, \bibinfo {author} {\bibfnamefont {Q.}~\bibnamefont {Turchette}},
  \bibinfo {author} {\bibfnamefont {M.}~\bibnamefont {Chapman}}, \ and\
  \bibinfo {author} {\bibfnamefont {H.}~\bibnamefont {Kimble}},\ }\href@noop {}
  {\bibfield  {journal} {\bibinfo  {journal} {Optics letters}\ }\textbf
  {\bibinfo {volume} {21}},\ \bibinfo {pages} {1393} (\bibinfo {year}
  {1996})}\BibitemShut {NoStop}%
\bibitem [{\citenamefont {Maitre}\ \emph {et~al.}(1997)\citenamefont {Maitre},
  \citenamefont {Hagley}, \citenamefont {Nogues}, \citenamefont {Wunderlich},
  \citenamefont {Goy}, \citenamefont {Brune}, \citenamefont {Raimond},\ and\
  \citenamefont {Haroche}}]{maitre1997quantum}%
  \BibitemOpen
  \bibfield  {author} {\bibinfo {author} {\bibfnamefont {X.}~\bibnamefont
  {Maitre}}, \bibinfo {author} {\bibfnamefont {E.}~\bibnamefont {Hagley}},
  \bibinfo {author} {\bibfnamefont {G.}~\bibnamefont {Nogues}}, \bibinfo
  {author} {\bibfnamefont {C.}~\bibnamefont {Wunderlich}}, \bibinfo {author}
  {\bibfnamefont {P.}~\bibnamefont {Goy}}, \bibinfo {author} {\bibfnamefont
  {M.}~\bibnamefont {Brune}}, \bibinfo {author} {\bibfnamefont
  {J.}~\bibnamefont {Raimond}}, \ and\ \bibinfo {author} {\bibfnamefont
  {S.}~\bibnamefont {Haroche}},\ }\href@noop {} {\bibfield  {journal} {\bibinfo
   {journal} {Physical review letters}\ }\textbf {\bibinfo {volume} {79}},\
  \bibinfo {pages} {769} (\bibinfo {year} {1997})}\BibitemShut {NoStop}%
\bibitem [{\citenamefont {Kumar}\ \emph
  {et~al.}(2022{\natexlab{b}})\citenamefont {Kumar}, \citenamefont
  {Suleymanzade}, \citenamefont {Stone}, \citenamefont {Taneja}, \citenamefont
  {Anferov}, \citenamefont {Schuster},\ and\ \citenamefont
  {Simon}}]{https://doi.org/10.48550/arxiv.2207.10121}%
  \BibitemOpen
  \bibfield  {author} {\bibinfo {author} {\bibfnamefont {A.}~\bibnamefont
  {Kumar}}, \bibinfo {author} {\bibfnamefont {A.}~\bibnamefont {Suleymanzade}},
  \bibinfo {author} {\bibfnamefont {M.}~\bibnamefont {Stone}}, \bibinfo
  {author} {\bibfnamefont {L.}~\bibnamefont {Taneja}}, \bibinfo {author}
  {\bibfnamefont {A.}~\bibnamefont {Anferov}}, \bibinfo {author} {\bibfnamefont
  {D.~I.}\ \bibnamefont {Schuster}}, \ and\ \bibinfo {author} {\bibfnamefont
  {J.}~\bibnamefont {Simon}},\ }\href {\doibase 10.48550/ARXIV.2207.10121}
  {\enquote {\bibinfo {title} {Quantum-limited millimeter wave to optical
  transduction},}\ } (\bibinfo {year} {2022}{\natexlab{b}})\BibitemShut
  {NoStop}%
\end{thebibliography}%

\end{document}